\documentclass[prb,twocolumn,showpacs,amsmath,amssymb,floats,floatfix,superscriptaddress]{revtex4}
\usepackage{graphicx}
\usepackage{amsmath}
\usepackage[dvips]{color}
\graphicspath{{eps/},{stm/}}
\begin{document}
\title{First-principles and semi-empirical van der Waals study of thymine on Cu(110) surface}
\author{V.~Caciuc}\email{v.caciuc@fz-juelich.de}
\affiliation{Physikalisches Institut, Westf\"alische Wilhelms Universit\"at
             M\"unster, Wilhelm-Klemm-Str.~10, 48149 M\"unster, Germany}
\affiliation{Institut f\"ur Festk\"orperforschung (IFF),
Forschungszentrum J\"ulich, 52425 J\"ulich, Germany}
\author{N.~Atodiresei}\email{n.atodiresei@fz-juelich.de}
\affiliation{Institut f\"ur Festk\"orperforschung (IFF),
Forschungszentrum J\"ulich, 52425 J\"ulich, Germany}
\affiliation{The Institute of Scientific and Industrial Research, 
Osaka University, 8-1 Mihogaoka, Ibaraki Osaka, 567-0047 Japan}
\author{P.~Lazi\'{c}\footnote{On leave of absence from the Rudjer Bo\v{s}kovi\'{c} Institute, Zagreb, Croatia.}}
\affiliation{Institut f\"ur Festk\"orperforschung (IFF),
Forschungszentrum J\"ulich, 52425 J\"ulich, Germany}
\author{Y.~Morikawa}
\affiliation{The Institute of Scientific and Industrial Research, 
Osaka University, 8-1 Mihogaoka, Ibaraki Osaka, 567-0047 Japan}
\author{S.~Bl\"ugel}
\affiliation{Institut f\"ur Festk\"orperforschung (IFF),
Forschungszentrum J\"ulich, 52425 J\"ulich,  Germany}
\date{\today}
\begin{abstract}

In this study we investigated by means of density functional theory
calculations the adsorption geometry and bonding mechanism of a single thymine
(C$_5$H$_6$N$_2$O$_2$) molecule on Cu(110) surface. In the most stable
energetic configuration, the molecular plane is oriented perpendicular to
substrate along the $[1\bar{1}0]$ direction. For this adsorption geometry, the
thymine molecule interacts with the surface via a deprotonated nitrogen atom
and two oxygen ones such that the bonding mechanism involves a strong
hybridization between the highest occupied molecular orbitals (HOMOs) and the
$d$-states of the substrate. In the case of a parallel adsorption geometry, 
the long-range van der Waals interactions play an important role on both the
molecule-surface geometry and adsorption energy. Their specific role was
analyzed by means of a semi-empirical and the seamless methods. In particular,
for a planar configuration, the inclusion of the dispersion effects
dramatically changes the character of the adsorption process from physisorption
to chemisorption. Finally, we predict the real-space topography of the
molecule-surface interface by simulating scanning tunneling microscopy (STM)
images. From these simulations we anticipate that only certain adsorption
geometries can be imaged in STM experiments.

\end{abstract}
\pacs{
  71.15.Dx  
  68.43.-h  
  68.43.Bc  
  68.43.Fg  
}
\maketitle
\section{Introduction}
\label{sec:intro}

The development of electronic devices like organic light-emitting diodes
(OLEDs)\cite{APL51_913,Shinar_bookOLED} has opened the road to an alternative
technology to the silicon-based one to manufacture integrated electronic
devices built on single molecules. Nowadays, the emerging field of molecular
electronics\cite{Science278_252,Nature408_541} focuses on the possibility of
using organic molecules adsorbed on surfaces as basic functional units to
construct technological relevant electronic devices such as single molecule
diodes,\cite{PNAS102_8815,JACS124_11862,PRL96_096803} organic field-effect
transistors\cite{Science283_822,AdvMat14_99,Science303_1644} or
ultra-high-density memory circuits.\cite{Nature445_414}

A very appealing feature of molecular devices is the prospect of tuning their
physical properties at atomic scale by chemical functionalization of the
molecule-surface interface under consideration. A key aspect of this approach
is to employ molecules with several functional groups that adsorb on surface
through one such group while the others can be used to react with other
molecules to selectively modify the geometry and electronic properties of
adsorbate-substrates systems. For instance, experimental studies aimed to
investigate the functionalization of the Cu(110) surface by adsorption of
organic molecules have been performed for terephthalic\cite{PRB66_155427} and
oxalic\cite{SS539_171} acids.

In this context, the chemical functionalization of surfaces via biologically
relevant molecules is of particular interest due to the potentially important
applications such as molecular biosensors.\cite{BB14_599,JIEEESens5_744,
Nanotech18_424017} In particular, one of the most important molecules in biology
is thymine (C$_5$H$_6$N$_2$O$_2$). Its importance stems from the fact that it
is a fundamental nucleobase of the deoxyribonucleic acid (DNA) and encodes
genetic information by pairing with the adenine base. The adsorption of thymine
on surfaces is then of particular interest since it represents a possible
channel for anchoring of DNA on substrates. The relevance of this process is
also highlighted by the observation that the thymine base is absent in the case
of the ribonucleic acid (RNA).

From experimental point of view, the adsorption process of thymine on Cu(110)
surface was intensively investigated by means of reflection-adsorption infrared
spectroscopy (RAIRS),\cite{SS502_185,SS561_233} (soft-)X-ray photoelectron
spectroscopy (XPS),\cite{SS532_261,SS601_3611} near-edge X-Ray adsorption fine
structure (NEXAFS)\cite{SS532_261,SS601_3611} and photo-electron
diffraction\cite{SS601_3611} at room or higher temperatures. These studies
suggest that, in the limit of low coverage, the geometry of the thymine-Cu(110)
system corresponds to an upright orientation of the molecular plane aligned
along $[1\bar{1}0]$ surface direction. However, the chemical identity of the 
functional groups involved in the bonding with the metal substrate is less
clear. One possibility is to consider that the thymine molecule binds to Cu(110)
surface via a (dehydrogenated) N atom adjacent to two carbonyl groups (CO)
while another possible adsorption geometry implies a adsorbate-substrate
interaction through the imino (CNH) and carbonyl (CO) groups.

In consequence, in the present study we focus on an \textit{ab initio}
investigation of the adsorption geometry and bonding mechanism of a single
thymine molecule on Cu(110) surface. The bonding geometry of the adsorbed
molecule was analyzed as a function of the orientation of the molecular plane
with respect to substrate. More specifically, we considered the case when the 
heterocyclic ring is perpendicular to surface and aligned along its $[001]$
and $[1\bar{1}0]$ directions. We also examined a parallel adsorption geometry
such that the molecular plane lies flat on surface. In particular, for this
bonding geometry the long-range van der Waals interactions are important
from geometrical and energetic point of view. The effect of the dispersion
corrections on the adsorption geometry was taken into account by means of a
semi-empirical approach while their impact on the adsorption energy was also
evaluated within the seamless (vdW-DF) method. Besides this, the bonding
mechanism of the thymine-Cu(110) system in the perpendicular adsorption
geometry was investigated considering a single and double deprotonated
C$_5$H$_6$N$_2$O$_2$ at N sites. In each case, we considered two scenarios for
the bonding functional groups as already discussed above. The experimental
fingerprints of all adsorption geometries analyzed in our study as revealed by
scanning tunneling microscope (STM) was also simulated.

\section{Theoretical method}
\label{sec:calc}

Our \textit{ab initio} study was performed within the framework of density
functional theory (DFT)\cite{PR136_B864} by using the generalized gradient
approximation (GGA) for the exchange-correlation energy
functional in the form proposed by Perdew-Burke-Ernzerhof
(PBE).\cite{PRL77_3865} The Kohn-Sham
equations\cite{PR140_A1133} have been solved self-consistently using the
pseudopotential method\cite{RMP64_1045} as implemented in the VASP
program.\cite{PRB47_558,PRB54_11169} The electron-ion interactions
are replaced by pseudopotentials described by the projector augmented-wave
method (PAW).\cite{PRB50_17953}

The thymine--Cu(110) system was modeled by a periodic slab geometry using
the theoretical lattice parameter of the bulk Cu (3.64\,{\AA}). In each
supercell, the slab consists of five atomic layers separated by a vacuum region
of $\approx$ 21\,{\AA}. The size of the in-plane surface unit cell was set to
5$\times$4. For this geometrical setup, the Kohn-Sham orbitals were expanded
over a plane-wave basis set which includes all plane waves up to a cutoff
energy E$_{cut}$ of 450\,eV. The geometry of the thymine--Cu(110) system was
optimized by relaxing the atomic positions of all molecule atoms and those in
two surface layers. The equilibrium geometry of the molecule-Cu(110) surface
was obtained when the calculated Hellmann-Feynman forces were smaller than
$\approx$ 0.001 eV/{\AA}. Finally, the Brillouin zone integrations were carried
out using only the $\Gamma$-point.

To investigate the structural stability of different conformational geometries
of a single or double deprotonated adsorbate-surface systems, we calculated the
adsorption energy $E_{ads}$ and adsorption enthalpy $\Delta H_{T=0}^{ads}$ at
a temperature T of 0 K for each geometry under consideration.

The adsorption energy $E_{ads}$ is given by
\begin{equation}
 E_{ads} =  E_{sys} - E_{Cu(110)} - E_{molecule}.
\label{eq:ads_ene}
\end{equation} 
where $E_{sys}$ represents the total energy of the adsorbed thymine-Cu(110)
surface  system, $E_{Cu(110)}$ is energy of the isolated Cu(110) surface and
$E_{molecule}$ denotes the energies of the isolated thymine molecular species
as follows: (i) $E_{Thymine}$ $-$thymine and (ii) $E_{Thymine}^{H,H_{2}}$
$-$single (superscript H ) or $-$double (superscript H$_{2}$) deprotonated
thymine molecule, respectively. 

The adsorption enthalpy $\Delta H_{T=0}^{ads}$ is defined as the difference
between the sum of the product energies ($E_{sys}$ and $E_{H_{2}}$ $-$the total
energy of the isolated hydrogen molecule $H_{2}$) and the sum of reactant
energies ($E_{Cu(110)}$ and $E_{Thymine}$). Therefore, the adsorption enthalpy 
$\Delta H_{T=0}^{ads}$ is defined by
\begin{align}
\begin{split}
 \Delta H_{T=0}^{ads} = 
         &( E_{sys} + factor*E_{H_{2}}) - \\
         &( E_{Thymine}+E_{Cu(110)}).
\label{eq:ads_enth1}
\end{split}
\end{align}
where $factor$ is equal 0.0 (for thymine molecule), 0.5 (for a single
deprotonated case) or to 1.0 (for a double deprotonated thymine molecule). A
negative adsorption enthalpy indicates an exothermic reaction between the
thymine molecule and the Cu(110) surface while a positive value of
$\Delta H_{T=0}^{ads}$ implies that such reaction is endothermic and requires
external energy to make it possible.

However, the double deprotonation process of the C$_5$H$_6$N$_2$O$_2$ molecule
can take place through an intermediate step involving a single deprotonated
thymine molecule. Two different scenarios are conceivable: (i) the single 
deprotonated molecule adsorbed on the Cu(110) surface loses one of its proton
such that the adsorption enthalpy is given by
\begin{align}
 \Delta H_{T=0}^{ads} = (E_{sys}^{H_2}+0.5*E_{H_{2}})-E_{sys}^{H}.
\label{eq:ads_enth2}
\end{align}
and (ii) a single deprotonated thymine species in gas phase adsorbs on 
the metal substrate where undergoes a second deprotonation such that
\begin{align}
\begin{split}
 \Delta H_{T=0}^{ads} = 
         &( E_{sys}^{H_2} + 0.5*E_{H_{2}}) - \\
         &( E_{Thymine}^{H}+E_{Cu(110)}).
\label{eq:ads_enth3}
\end{split}
\end{align}
Thus, the adsorption enthalpy $\Delta H_{T=0}^{ads}$ calculated for a double
deprotonated C$_5$H$_6$N$_2$O$_2$ molecule by Eqs.~\ref{eq:ads_enth1},
~\ref{eq:ads_enth2} and ~\ref{eq:ads_enth3} will allow us to get a hint on
the details of the deprotonation process of this molecule on Cu(110) surface.

An important question arising when studying the adsorption process of
molecules on metal surfaces concerns the role of long-range van der Waals
interactions on the geometry and the adsorption energy of the
adsorbate-substrate system in question. We addressed this problem in our study
by investigating the thymine molecule on Cu(110) surface also in the framework
of the semi-empirical approach (DFT-D) as proposed by Grimme.\cite{JCC27_1787}
The basic advantage of this method is the possibility to add the self-consistent
calculated van der Waals forces to the Hellmann-Feynman ones to fully relax
the molecule-surface system. We previously emphasized the importance of the
inclusion of these dispersion forces to obtain a reliable adsorbate-substrate
geometry in the case of the pyridine adsorbed on Cu(110) and Ag(110)
surfaces\cite{PRB78_045411} or in the case of a N,N${\prime}$-di($n$-butyl)
quinacridone monolayer on the Ag(110) substrate.\cite{PRB78_165432} A similar
semi-empirical approach was used to study the interaction of adenine with
graphite(0001) surface\cite{PRL95_186101} or applied to noble-gas, N$_2$ and
benzene dimers.\cite{PRB73_205101} Additionally, the energetics of the
thymine-Cu(110) substrate was analyzed by means of the seamless approach
(vdW-DF) proposed by Dion\textit{et al}.\cite{PRL92_246401} In this
\textit{ab initio} method, the vdW interactions are evaluated in a
non-self-consistent procedure as a nonlocal correlation energy functional.
This approach was already employed to investigate, for instance, benzene,
naphthalene and phenol on graphite(0001),\cite{PRL96_146107,PRB74_155402} or
thiophene on Cu(110) surfaces.\cite{PRL99_176401}

\section{Results and discussion}
\label{sec:results}

\subsection{Thymine in gas phase}
\label{sec:thym}

To analyze how the molecule-substrate interaction modifies the geometry of the
adsorbed thymine (C$_5$H$_6$N$_2$O$_2$) on Cu(110) surface, we first optimized
the geometry of this molecule in the gas phase [see
Fig.~\ref{fig:Thym}(a)]. The structural optimization was performed by using a
cubic unit cell of 20$\times$20$\times$20{\AA}$^3$. The calculated bond
lengths and bond angles are reported in Table~\ref{tab:BondLength}. For
comparison, in this table we also list the data of similar DFT calculations
performed in Ref.~\onlinecite{IntJMolSci1_17}. Overall, one can observe that
both studies provide a similar description of the molecular geometry.
\begin{table*}[htb]
\center{
\begin{tabular}{*{6}{c@{\quad}}}
\hline\hline
           & \multicolumn{2}{c}{Gas phase} & \multicolumn{3}{c}{Cu(110)} \\
	   & this work  & Ref.~\onlinecite{IntJMolSci1_17}
	   &  &  & \\
 Configuration
           &               &                
	   & Perpendicular & \multicolumn{2}{c}{Parallel} \\
 Orientation
	   &               &
	   & $[1\bar{1}0]$ & $[1\bar{1}0]$ & $[001]$      \\
	   &               &
	   & DFT & \multicolumn{2}{c}{DFT+vdW} \\
\hline
\multicolumn{6}{c} {Bond Length ({\AA})}  \\
 C1-O1     & 1.232 & 1.222 & 1.266 & 1.232 & 1.229 \\
 C2-O2     & 1.229 & 1.218 & 1.262 & 1.248 & 1.251 \\
 C1-N1     & 1.412 & 1.408 & 1.376 & 1.427 & 1.430 \\
 C2-N1     & 1.385 & 1.386 & 1.356 & 1.372 & 1.369 \\
 C2-N2     & 1.392 & 1.390 & 1.379 & 1.375 & 1.370 \\
 C3-N2     & 1.375 & 1.380 & 1.367 & 1.378 & 1.379 \\
 C3-C4     & 1.358 & 1.352 & 1.360 & 1.363 & 1.362 \\
 C4-C5     & 1.511 & 1.501 & 1.511 & 1.508 & 1.506 \\
 C4-C1     & 1.479 & 1.468 & 1.468 & 1.473 & 1.475 \\
 N-Cu      &       &       & 1.988 &       &       \\
 O1-Cu     &       &       & 2.017 &       &       \\
 O2-Cu     &       &       & 2.028 & 2.218 & 2.135 \\
\multicolumn{6}{c} {Bond Angle ({$^\circ$})}  \\
 O1-C1-N1  & 119.9 & 120.4 & 120.8 & 119.6 & 119.6 \\
 O2-C2-N1  & 124.5 &       & 124.8 & 123.8 & 122.9 \\
 C1-N1-C2  & 128.6 & 128.3 & 121.6 & 127.8 & 127.6 \\
 C2-N2-C3  & 124.2 & 123.9 & 122.8 & 123.1 & 123.2 \\
 N2-C3-C4  & 122.9 & 122.7 & 121.1 & 123.2 & 122.8 \\
 C3-C4-C5  & 123.8 & 124.1 & 123.1 & 123.5 & 123.8 \\
 C3-C4-C1  & 117.8 & 118.2 & 116.7 & 117.7 & 118.0 \\
 C4-C1-O1  & 125.9 &       & 119.5 & 126.5 & 126.5 \\
 C4-C1-N1  & 114.2 & 114.5 & 119.7 & 113.9 & 113.8 \\
\hline\hline
\end{tabular}
}
\caption{The bond lengths and bond angles calculated for the thymine molecule
in gas phase and adsorbed on Cu(110) surface. In the case of a perpendicular
adsorption geometry, only the values obtained for the ground state
[see Fig.~\ref{fig:Thym}(b)] are presented. In the case of a parallel adsorption
geometry, only the data of the most stable configurations are reported. In
particular, the geometry corresponding to the orientation $[001]$ is sketched
in Fig.~\ref{fig:Thym}(d). In the orientation $[1\bar{1}0]$, the O1--O2 ''bond''
line is oriented parallel to this surface direction. The atomic labels are
those indicated in Fig.~\ref{fig:Thym}(a).}
\label{tab:BondLength}
\end{table*}

Similarly, the first step in understanding the electronic structure of the
thymine-surface interface is to analyze the electronic structure of the
C$_5$H$_6$N$_2$O$_2$ molecule in the gas phase. Therefore, in
Fig.~\ref{fig:LDOS}(a) we present the corresponding angular-momentum resolved
local density of states (LDOS). The highest occupied
molecular orbital (HOMO) as well as the lowest unoccupied molecular orbital
(LUMO) are $\pi$-like molecular orbitals, originating mainly from the atomic
$p$ orbitals perpendicular on the molecular plane (pyrimidine ring, consisting
of C and N atoms). The molecular orbital HOMO-1 and HOMO-2 have a $\sigma$
character indicating that the chemical bonding takes place in the molecular
plane.

\subsection{Ground-state geometry and energetics}
\label{sec:geom}

\subsubsection{Perpendicular adsorption geometry}

\begin{figure}[tb]
\begin{center}
  \begin{tabular}{cc}
  \multicolumn{2}{c}{\includegraphics[width=0.3\linewidth]{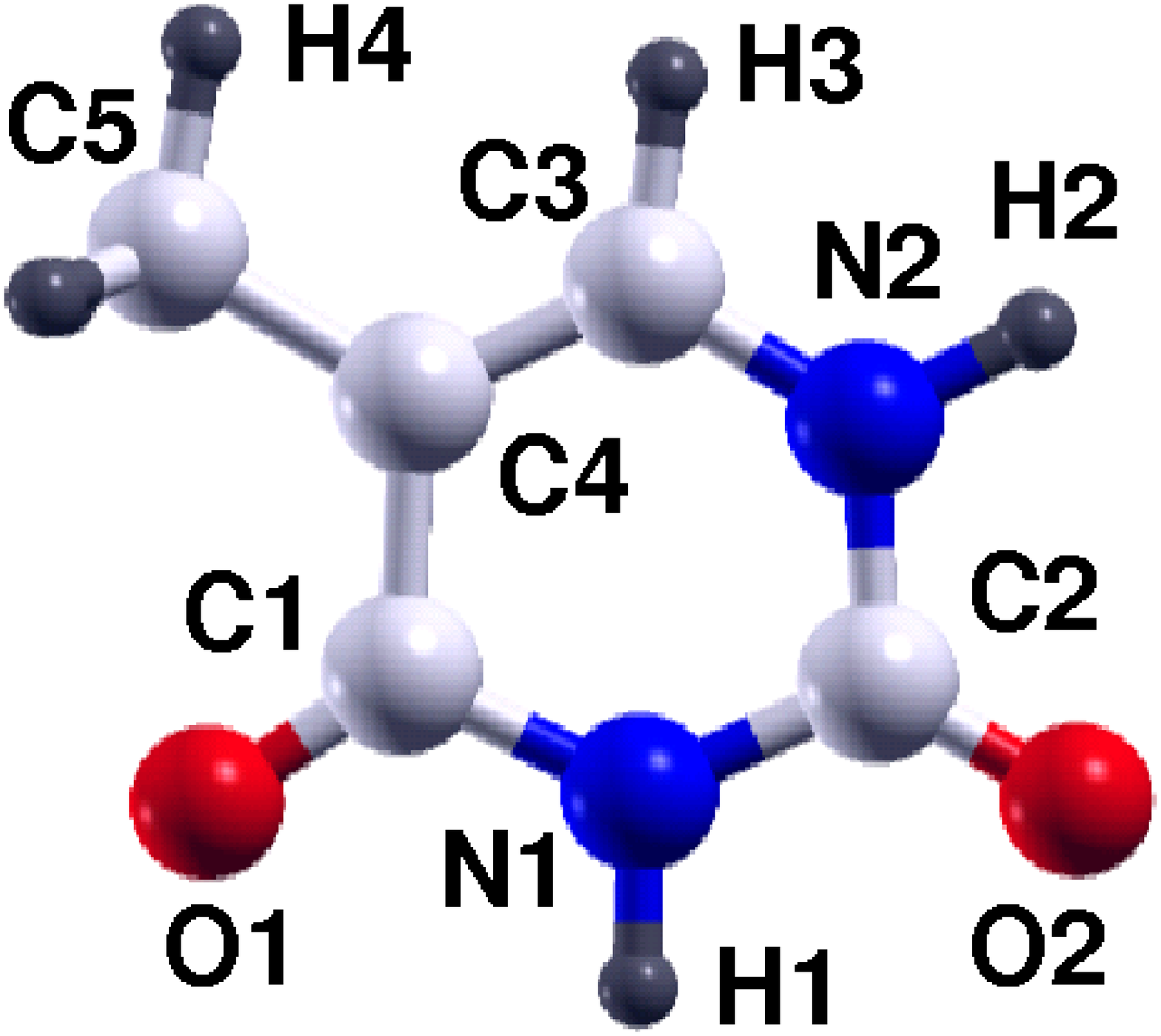}} \\
  \multicolumn{2}{c}{\textbf{(a)}} \\
  {\includegraphics[width=0.4\linewidth]{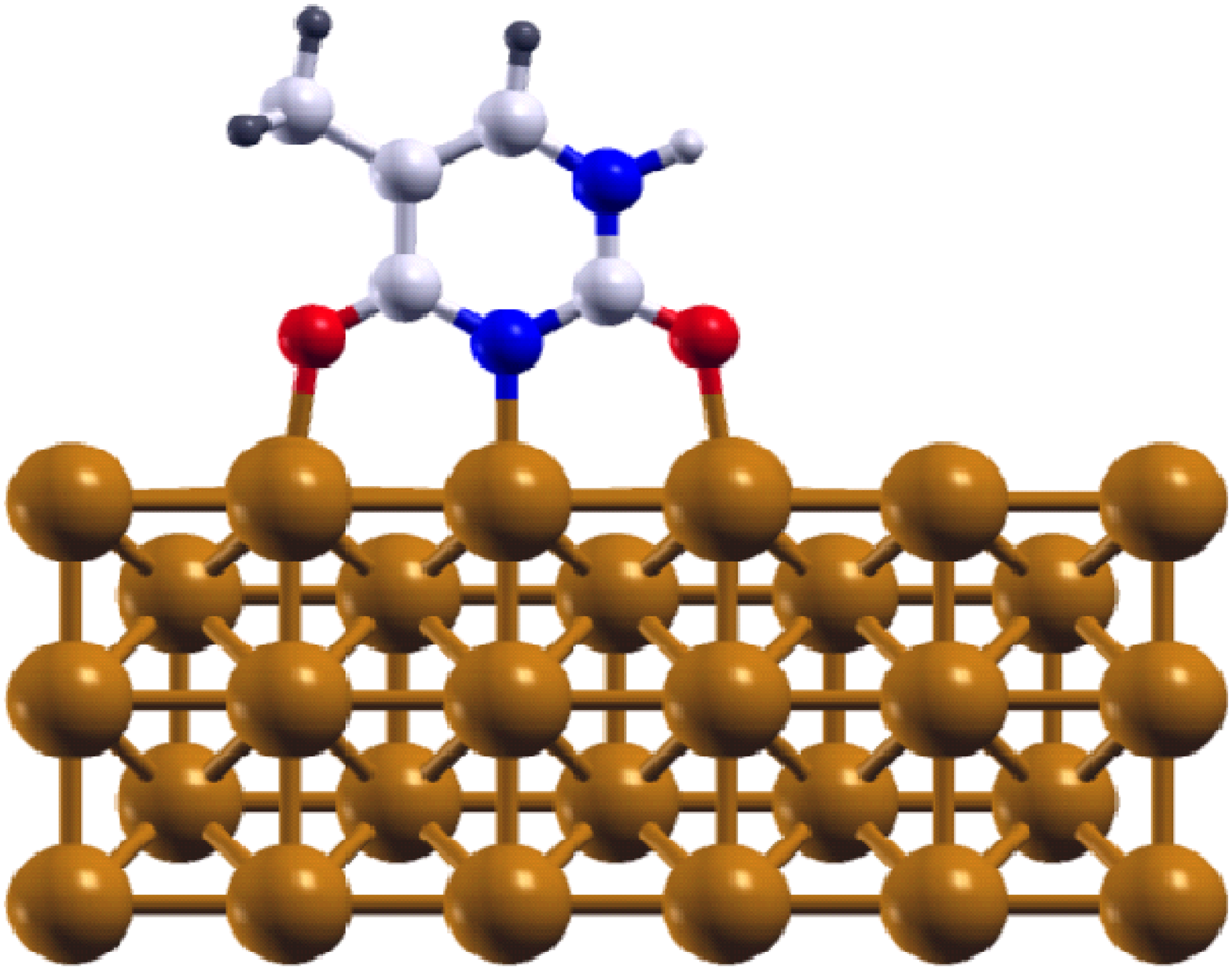}} &
  {\includegraphics[width=0.4\linewidth]{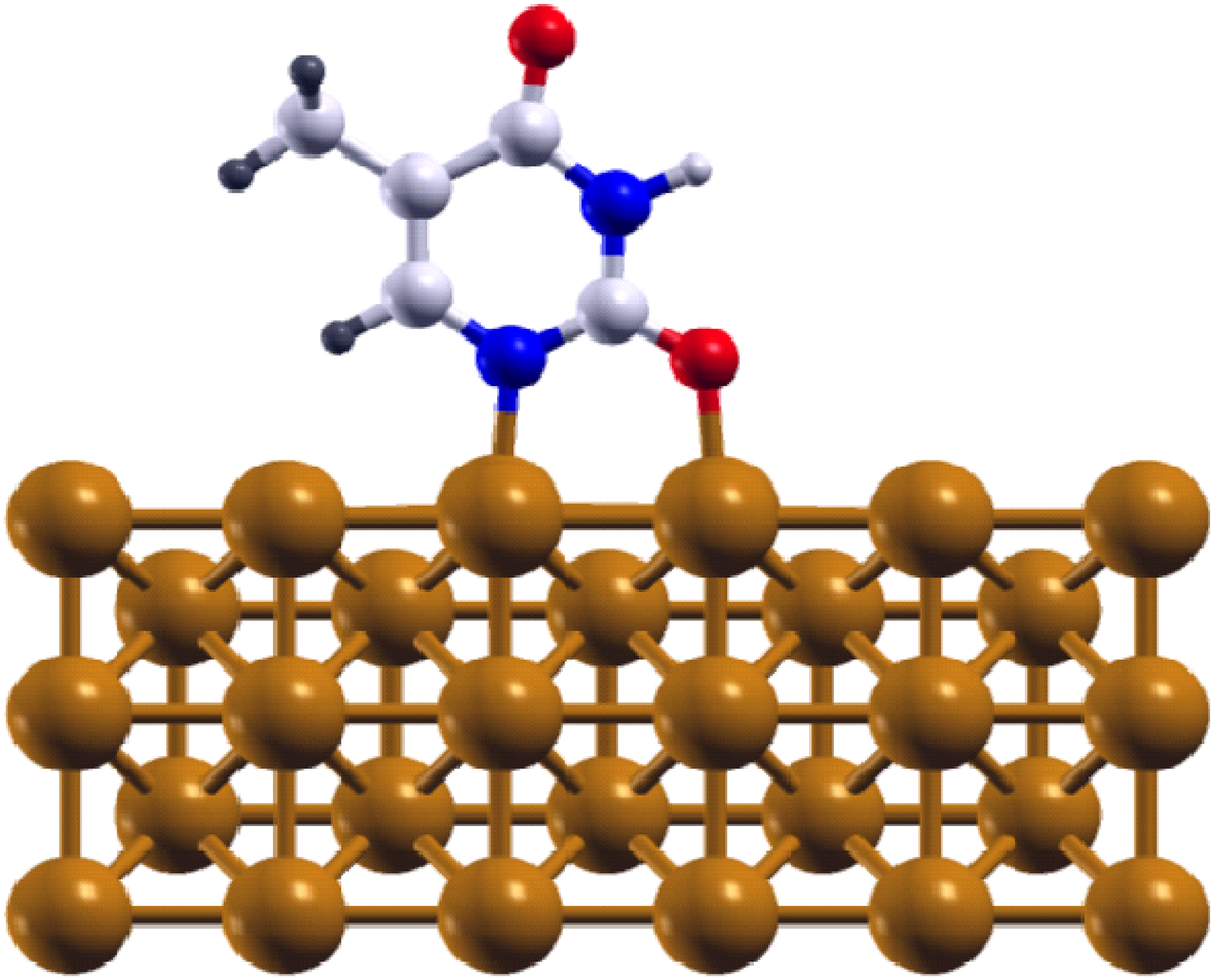}}  \\
  {\textbf{(b)}} & {\textbf{(c)}} \\
  {\includegraphics[width=0.4\linewidth]{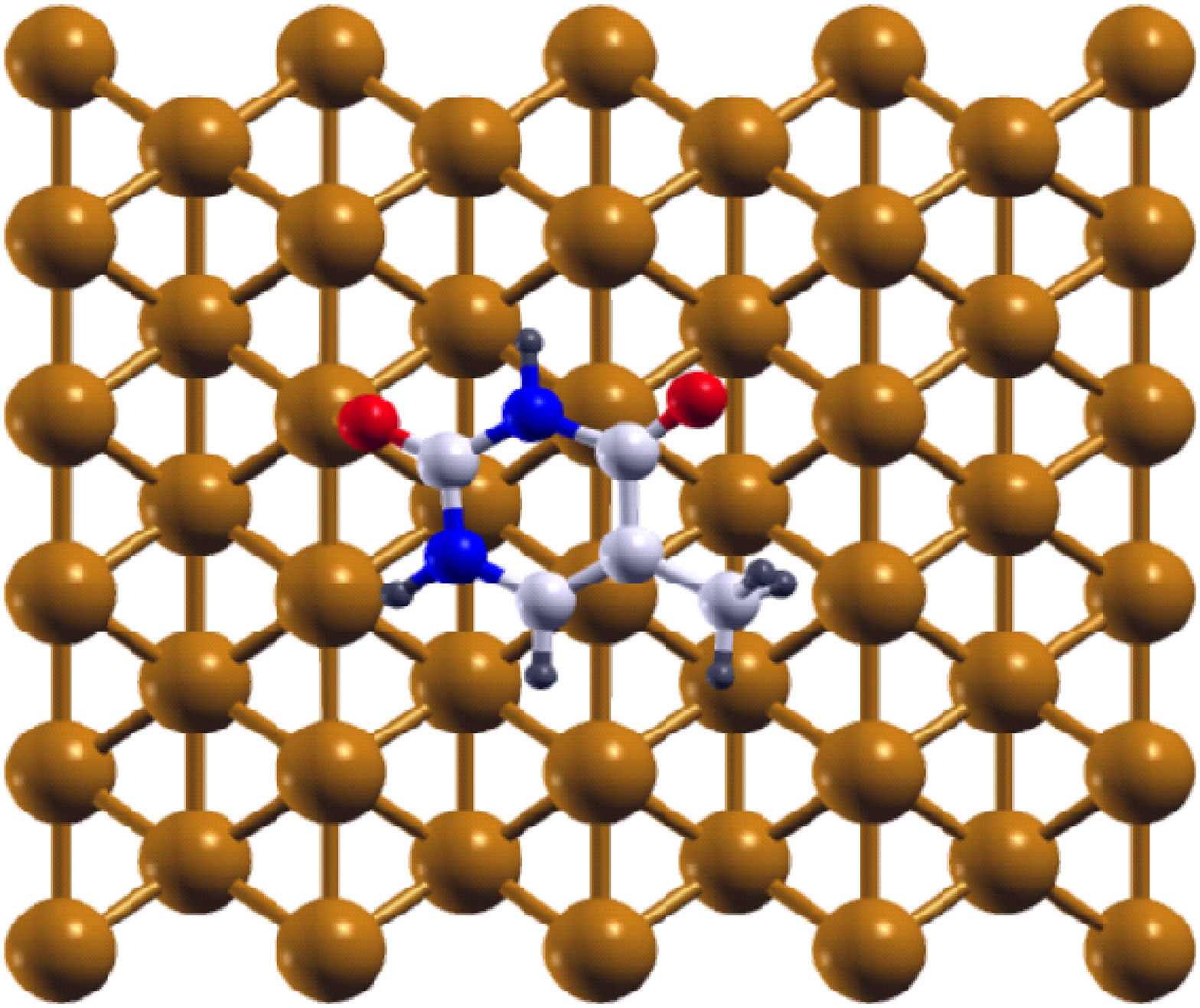}} &
  {\includegraphics[width=0.4\linewidth]{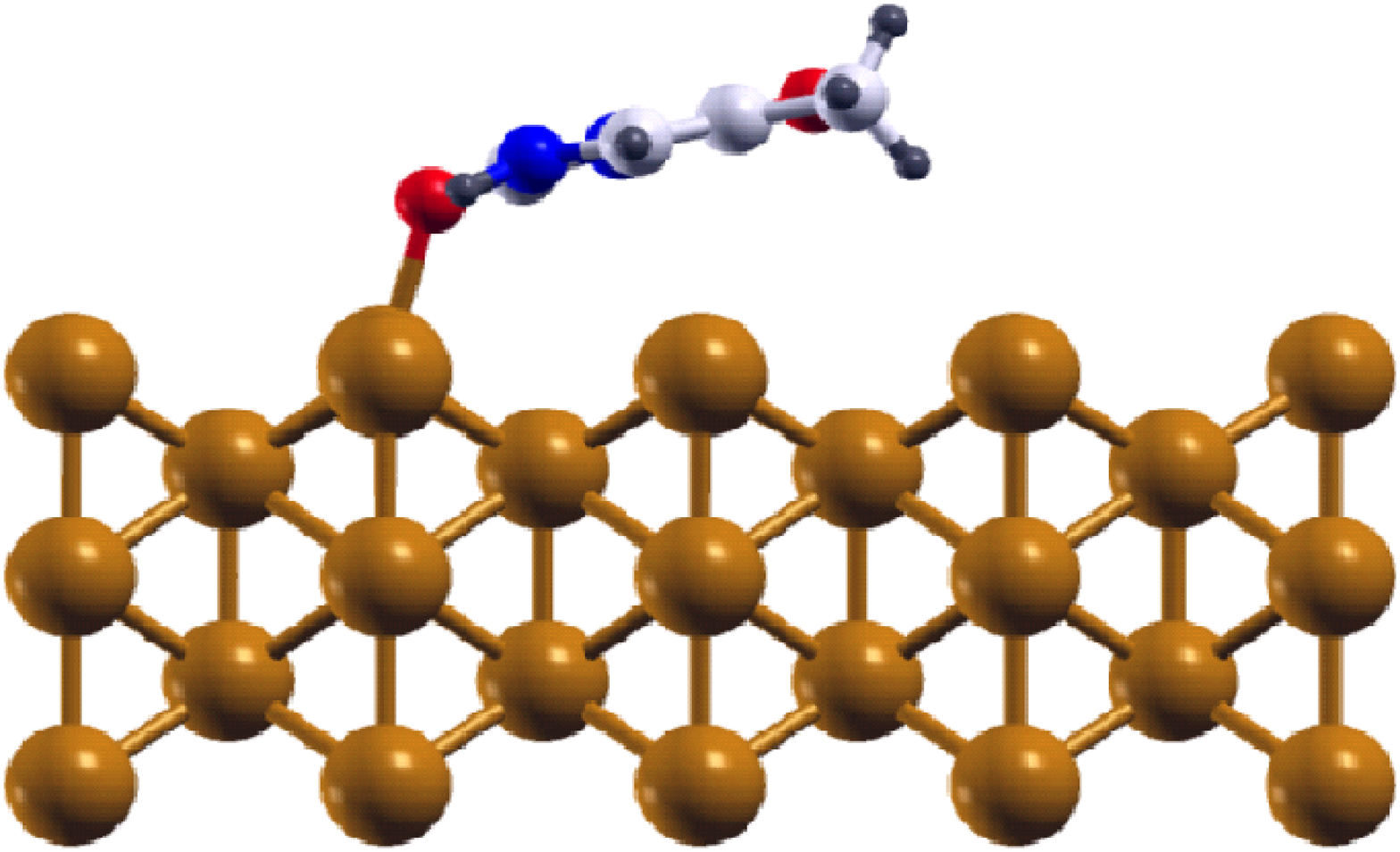}}  \\
  {\textbf{(d)}} & {\textbf{(e)}} \\
  \end{tabular}
\end{center}
\caption{
\textbf{(a)} Ball-and-stick model of the thymine molecule in the gas phase.
\textbf{(b)} In the ground-state, the thymine molecule adsorbs on the Cu(110)
surface via three bonds (N1--Cu,O1--Cu,O2--Cu) with the molecular plane
perpendicular to surface and oriented along the $[1\bar{1}0]$ surface
direction. This adsorption configuration is labeled as ONO--H in
Table~\ref{tab:ads_ene}. An additional adsorption geometry obtained by
removing the hydrogen atom colored in grey (H2 in \textbf{(a)}) was also
investigated (denoted as the geometry ONO--H$_2$ in Table~\ref{tab:ads_ene}).
\textbf{(c)} Alternatively, the thymine can attach to Cu(110) surface via 
two chemical bonds (N2--Cu and O1--Cu). For this configuration one can
consider also the cases of a single (NO--H geometry in Table~\ref{tab:ads_ene})
or a double (NO--H$_2$ in Table~\ref{tab:ads_ene}) deprotonated thymine
molecule.
\textbf{(d)} Top and \textbf{(e)} lateral view of the most stable
configuration of a parallel geometry of thymine on Cu(110). The O1--O2 ''bond''
line is oriented parallel to $[001]$ surface direction. The plots presented in
the present work have been obtained by using XCrysDen.\cite{XCrySDen}
}
\label{fig:Thym}
\end{figure}

\paragraph*{Without vdW interactions} --
To explore the possibility to chemically functionalize the Cu(110) surface
via the adsorption of a single thymine molecule on it, in our study we
investigated several perpendicular adsorbate-substrate geometries. In all
these adsorption configurations, the thymine binds to surface via the N atom
after a deprotonation process involving a N--H bond breaking. A binding of
thymine on the Cu(110) substrate through a N atom is similar to the pyridine
on Cu(110) and Ag(110) surfaces.\cite{PRB78_045411}

However, in the case of the thymine-Cu(110) system, there
are two different possible bonding N atoms denoted as N1 and N2 in
Fig.~\ref{fig:Thym}(a). The molecule-surface geometry corresponding to the
adsorption through the N1 atom is presented in Fig.~\ref{fig:Thym}(b) 
(hereafter referred to as ONO--H configuration) and that corresponding to the
bonding via the N2 atom is depicted in Fig.~\ref{fig:Thym}(c) (hereafter
referred to as NO--H configuration). In both cases, the molecular
plane given by the pyrimidine ring is oriented along the $[1\bar{1}0]$ surface
direction. In the present work we also investigated the case when the molecular
plane is parallel to the $[001]$ surface direction. Additionally, for each
adsorption geometry already mentioned, we considered the possibility that the
C$_5$H$_6$N$_2$O$_2$ molecule on Cu(110) surface is double deprotonated, i.e., 
both N--H bonds are broken.
\begin{table*}[htb]
\center{
\begin{tabular}{*{12}{c}}
\hline\hline
               &   & \multicolumn{4}{c}{adsorption energy}
                   & \multicolumn{6}{c}{adsorption enthalpy}         \\
               &   & \multicolumn{4}{c}{$E_{ads}$ (eV)}
	           & \multicolumn{6}{c}{$\Delta H_{T=0}^{ads}$ (eV)} \\
geometry       & orientation
	           & DFT(PBE) & DFT-D &         \multicolumn{2}{c}{vdW-DF}
		                              & \multicolumn{3}{c}{DFT(PBE)}
		                              & \multicolumn{3}{c}{DFT-D} \\
               &   & & & PBE & revPBE & & & & & & \\
               &   & & & & & Eq.~\ref{eq:ads_enth1}
	                   & Eq.~\ref{eq:ads_enth2}
			   & Eq.~\ref{eq:ads_enth3}
	                   & Eq.~\ref{eq:ads_enth1}
	                   & Eq.~\ref{eq:ads_enth2}
			   & Eq.~\ref{eq:ads_enth3} \\
\hline
\multicolumn{12}{c}{Perpendicular} \\
ONO-H     & $\begin{array}{c}{[1\bar{1}0]} \\ {[001]} \end{array}$
          & $\begin{array}{c}{-3.666} \\ {-3.027} \end{array}$
          & $\begin{array}{c}{-4.011} \\ {-3.396} \end{array}$
	  & $\begin{array}{c}{-4.078} \\ {-3.566} \end{array}$	  
	  & $\begin{array}{c}{-3.613} \\ {-3.157} \end{array}$
          & $\begin{array}{c}{-0.618} \\ {+0.022} \end{array}$
	  &  &
          & $\begin{array}{c}{-0.621} \\ {-0.006} \end{array}$
	  &  &  \\
ONO-H$_2$ & $\begin{array}{c}{[1\bar{1}0]} \\ {[001]} \end{array}$
          & $\begin{array}{c}{-4.643} \\ {-3.902} \end{array}$
          & $\begin{array}{c}{-4.994} \\ {-4.282} \end{array}$
	  & $\begin{array}{c}{-5.091} \\ {-4.438} \end{array}$
	  & $\begin{array}{c}{-4.623} \\ {-3.994} \end{array}$
	  & $\begin{array}{c}{+0.832} \\ {+1.573} \end{array}$
	  & $\begin{array}{c}{+1.450} \\ {+1.551} \end{array}$
	  & $\begin{array}{c}{-2.217} \\ {-1.475} \end{array}$
          & $\begin{array}{c}{+0.822} \\ {+1.535} \end{array}$
	  & $\begin{array}{c}{+1.443} \\ {+1.540} \end{array}$
	  & $\begin{array}{c}{-2.510} \\ {-1.800} \end{array}$ \\
 NO-H     & $\begin{array}{c}{[1\bar{1}0]} \\ {[001]} \end{array}$
          & $\begin{array}{c}{-2.674} \\ {-2.561} \end{array}$
          & $\begin{array}{c}{-3.050} \\ {-2.960} \end{array}$
	  & $\begin{array}{c}{-3.042} \\ {-2.987} \end{array}$
	  & $\begin{array}{c}{-2.668} \\ {-2.610} \end{array}$
	  & $\begin{array}{c}{-0.064} \\ {+0.048} \end{array}$
	  &  &
          & $\begin{array}{c}{-0.098} \\ {-0.007} \end{array}$
	  &  &  \\
 NO-H$_2$ & $\begin{array}{c}{[1\bar{1}0]} \\ {[001]} \end{array}$
          & $\begin{array}{c}{-3.160} \\ {-3.136} \end{array}$
          & $\begin{array}{c}{-3.590} \\ {-3.510} \end{array}$
	  & $\begin{array}{c}{-3.570} \\ {-3.591} \end{array}$
	  & $\begin{array}{c}{-3.198} \\ {-3.208} \end{array}$
	  & $\begin{array}{c}{+2.316} \\ {+2.340} \end{array}$
	  & $\begin{array}{c}{+2.380} \\ {+2.292} \end{array}$
	  & $\begin{array}{c}{-0.294} \\ {-0.270} \end{array}$
          & $\begin{array}{c}{+2.228} \\ {+2.307} \end{array}$
	  & $\begin{array}{c}{+2.326} \\ {+2.314} \end{array}$
	  & $\begin{array}{c}{-0.578} \\ {-0.450} \end{array}$ \\
\multicolumn{12}{c}{Parallel} \\
 flat   & $\begin{array}{c}{[1\bar{1}0]} \\ {[001]} \end{array}$
        & $\begin{array}{c}{+0.296} \\ {+0.112} \end{array}$
        & $\begin{array}{c}{-0.260} \\ {-0.319} \end{array}$
	& $\begin{array}{c}{-0.516} \\ {-0.703} \end{array}$
	& $\begin{array}{c}{-0.002} \\ {-0.247} \end{array}$
        & $\begin{array}{c}{-}      \\ {-} \end{array}$
	&  &
        & $\begin{array}{c}{-}      \\ {-} \end{array}$
	&  &  \\
\hline\hline
\end{tabular}
}
\caption{The adsorption energies $E_{ads}$ and adsorption enthalpies
$\Delta H_{T=0}^{ads}$ for the thymine--Cu(110) adsorption geometries
considered in our study [see Fig.~\ref{fig:Thym}(b), (c), (d) and (e),
respectively]. Here DFT(PBE) denotes the first-principles DFT results obtained
with the PBE\cite{PRL77_3865} exchange-correlation energy functional, DFT-D
represents the DFT results plus the semiempirical vdW contribution as proposed
by Grimme\cite{JCC27_1787} and vdW-DF implies the DFT results obtained by using
the Dion\cite{PRL92_246401} functional. Note that in the case of the vdW-DF
calculations we employed PBE\cite{PRL77_3865} as well as revPBE\cite{PRL80_890}
exchange-correlation energy functionals.
}
\label{tab:ads_ene}
\end{table*}

To determine which adsorption configuration is energetically the most
favorable one, we compared the corresponding adsorption energies reported in
Table~\ref{tab:ads_ene}. One can easily observe that, for all perpendicular
thymine-surface configurations considered in our study, in the ground-state
adsorption geometry the molecular plane is oriented parallel to $[1\bar{1}0]$
direction. Also, a lower adsorption energies is obtained for the ONO--H
adsorption configuration than for the NO--H one and a similar conclusion is
drawn for the ONO--H$_2$ geometry when compared with the NO--H$_2$ one. This
behavior is not surprising since early studies of formate\cite{PRB75_115407}
and terephthalic acid\cite{PRB76_115433} on Cu(110) surface revealed that the
oxygen atoms of these molecules adsorb on top of the surface atoms. As clearly
shown in Fig.~\ref{fig:Thym}(b) and (c), in the case of the $[1\bar{1}0]$
adsorption geometries, both thymine oxygen atoms lie above the surface Cu
ones. One the other hand, in the case of the ONO--H (ONO--H$_2$) configurations
there are two oxygen atoms above the surface one, in contrast to the case of
the NO--H (NO--H$_2$) geometries where only one oxygen atom interacts directly
with the surface.

Also, the calculated adsorption energies for the geometries with a
double deprotonated thymine on the Cu(110) substrate are lower than those
obtained for the single deprotonated adsorption configurations. However, since
for these adsorbate-surface geometries the number of atoms is different, it is
not possible to compare directly the corresponding adsorption energies to
infer which configuration is the most stable one. However, by comparing the
calculated adsorption enthalpies $\Delta H_{T=0}^{ads}$ one can determine if
the deprotonation process of thymine molecule is favorable from energetic
point of view or not. In particular, this information is of peculiar importance
for a double deprotonated thymine since in this case the $\Delta H_{T=0}^{ads}$
and the relative surface free energies $\Delta \gamma$ plotted as a function of
the chemical potential $\mu_H$ of the H atom will allow us to determine which
reaction path corresponding to Eq.~\ref{eq:ads_enth1}, Eq.~\ref{eq:ads_enth2}
or to Eq.~\ref{eq:ads_enth3} is the most probable one.

In the light of the above considerations, from Table~\ref{tab:ads_ene} one can
deduce that for a single deprotonated thymine molecule on Cu(110) surface
(ONO--H and NO--H configurations) the ground-state $[1\bar{1}0]$ adsorption
geometry does not require external energy to deprotonate the
C$_5$H$_6$N$_2$O$_2$ molecule, in contrast to the $[001]$ configuration.

In the case of a double deprotonated molecule, the large adsorption energy
which is a measure of the binding strength suggests that the double
deprotonated species might coexists together with single deprotonated species
on the metal surface. An important question is how a double deprotonated
species would arise. In our study we considered three different scenarios for
a double deprotonation process: 
 \newline (i) the reaction path suggested by Eq.~\ref{eq:ads_enth1} implies an 
instantaneous double deprotonation process when the thymine adsorbs on the 
Cu(110) surface. This could take place at high temperatures since the reaction 
is endoterm (requires external energy to take place);
 \newline (ii) the reaction path described by Eq.~\ref{eq:ads_enth2} implies that
the single deprotonated thymine adsorbed on the Cu(110) loses instantaneously
its second hydrogen yielding a double deprotonated species adsorbed on the
surface. This process is also an endothermic one and requires a larger amount 
of energy to take place as compared to (i); 
 \newline (iii) the Eq.~\ref{eq:ads_enth3} assumes a reaction path in which a single deprotonated
thymine molecule in gas phase adsorbs on the metal substrate where it undergoes
a second deprotonation process. Since a large amount of energy is released, the
adsorption process is an exotherm one and the probability to occur is can be
significant (depending on the hight of the energy barrier of the
transition state).

\begin{figure}
\begin{center}
  {\includegraphics[width=1.0\linewidth]{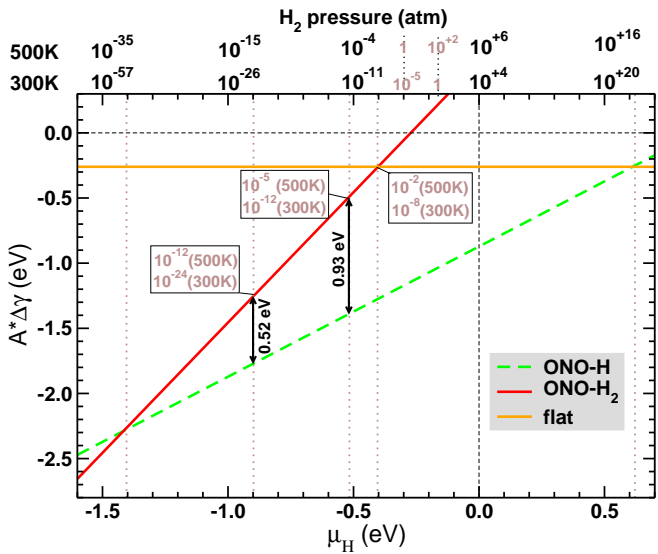}} \\
\end{center}
\caption{Relative surface free energies $\Delta \gamma$ of the flat,
ONO--H and ONO--H$_2$ configurations as function of the $H$ chemical potential
$\mu_{H}$. The values of the $\Delta \gamma$ are normalized
by the surface area $A$ of the unit cell (see Appendix). At the top of the
$x-$axis, the chemical potential of the $H_2$ molecule $\mu_{H_2}$ is given
as a pressure scale at two fixed temperatures: 300 K and 500 K.}
\label{fig:SurfEne}
\end{figure}

Since the adsorbed species of the flat thymine, ONO--H and ONO--H$_2$ differ
only through the number of hydrogen atoms, in Fig.~\ref{fig:SurfEne} we have
plotted the relative surface free energies $\Delta \gamma$ as a function of
the chemical potential $\mu_H$ of the H atom. Details of how $\Delta \gamma$
was calculated are presented in Appendix. At a given temperature, the chemical
potential of the H atom can be expressed in terms of the H$_2$ molecular
pressure which together with the temperature represent the parameters
used in experiments to control which conformation of the thymine (single or
double dehydrogenated) is adsorbed on the Cu(110) surface. The most stable
configuration is the one that minimizes the relative surface free energy
$\Delta \gamma$. In consequence, the Fig.~\ref{fig:SurfEne} indicates that,
independent on the temperature, the most stable configuration is ONO--H
for the chemical potentials of the H atom between $-0.9$ and $0.0$~eV.
Note that these values correspond to the chemical potential interval in which
the experiments can be performed. However, the
experiments\cite{SS532_261,SS561_233,SS502_185,SS601_3611} indicate that by
preparing the sample at $300$~K the ONO-H species are present on the surface
while when increasing the temperature higher than $500$~K the ONO-H$_2$
species are predominat at an experimental pressure of $10^{-12}$~atm. 
Based on the the experimental observations and as shown by our
\textit{ab initio} calculations (see Fig.~\ref{fig:SurfEne}) we conclude
that at room temperature ($\approx 300$~K) there is a high activation energy
between ONO--H and ONO--H$_2$ adsorbed species and therefore the second
deprotonation process cannot take place. Increasing the temperature over
$500$~K this energetic  barrier can be overcome such that the ONO-H$_2$
species are formed.\cite{SS601_3611}

To conclude, our first-principles calculations suggest that a double
deprotonated thymine molecule adsorbed on Cu(110) surface can be formed only
at high temperatures as follows: (i) the thymine adsorbs on the Cu(110)
surface undergoing an instantaneous double deprotonation process (see
Eq.~\ref{eq:ads_enth1})or (ii) the single deprotonated thymine
adsorbed on the Cu(110) loses its second hydrogen  yielding a double
deprotonated species adsorbed on the surface (process described by
Eq.~\ref{eq:ads_enth2}). Note that even if the adsorption enthalpy calculated
for the deprotonation process described by Eq.~\ref{eq:ads_enth3} is negative
and thus this process is exothermic, a single deprotonated pyridine molecule
in gas phase requires to be desorbed from the Cu(110) surface. In consequence,
to initialize this process one needs to provide an amount of externel energy
that equals at least the adsorption energy of the ONO-H or NO-H configurations
and therefore this scenario is less probable than those described by
Eqs.~\ref{eq:ads_enth1} and ~\ref{eq:ads_enth2}.

\paragraph*{Role of the vdW interactions} -- 
When including the van der Waals interactions using the DFT-D approach, all
relaxed thymine-Cu(110) geometries considered in our study remain practically
the same as those obtained without including the effect of the dispersion
corrections. This outcome of our \textit{ab initio} calculations is not
surprising since in all perpendicular adsorption configurations the thymine
molecule form a chemical bond with the surface via a N atom. As already
discussed by Thonhauser \textit{et al.},\cite{PRB76_125112} for covalently
bonded systems like a CO$_2$ molecule or the bulk silicon, the inclusion of
the vdW interactions does not alter the geometry obtained by DFT calculations.

However, the dispersion corrections will generally lower the adsorbtion energy
of an adsorbate-substrate system. The importance of this effect for chemisorbed
systems was emphasized in the study of the adsorption mechanism of benzene on
Si(001) surface performed by Johnston \textit{et al}.\cite{PRB77_121404(R)} 
Their first-principles calculations based on the vdW-DF
approach\cite{PRL92_246401} revealed that the PBE and revised PBE
(revPBE)\cite{PRL80_890} calculations predict a different ground-state
geometry than the vdW-DF ones. Therefore, to take into account the weak vdW
interactions might be important even for the strongly chemisorbed
molecule-surface systems. As presented in Table~\ref{tab:ads_ene}, in the case
of thymine-Cu(110) surface the $[1\bar{1}0]$ adsorption geometry remains the
ground-state one for all four adsorption configurations even in the presence
of the attractive van der Waals interactions. In the semi-empirical (DFT-D)
approach, their main effect is to lower the adsorption energy calculated for
the perpendicular adsorbate-surface geometries by about 0.3 eV with respect to
the values obtained with PBE. Note that this energy gain due to the dispersion
corrections is similar to that obtained for pyridine on Cu(110) and Ag(110)
surfaces.\cite{PRB78_045411} As regarding the adsorption enthalpies, their
values are also lowered when the dispersion effects are included. 

A final note of this section concerns the comparison of the adsorption energies
obtained with the semi-empirical (DFT-D) and seamless (vdW-DF) methods, the
latter values being obtained with a code developed by us.\cite{CPC} We briefly
remind that in the vdW-DF approach the correlation part of the GGA functional
is replaced by a sum of the LDA correlation functional $E_{LDA,c}$ and a
nonlocal term $E_{c}^{nl}$:
\begin{equation}
E_{vdW-DF}=E_{GGA}-E_{GGA,c}+E_{LDA,c}+E_{c}^{nl}.
\label{eq:vdW-DF}
\end{equation}
where $E_{GGA}$ is the self-consistent GGA total energy $E_{GGA}$ evaluated
in a conventional DFT calculation. The nonlocal correlation energy $E_{c}^{nl}$
can be expressed as a function of the charge density $n(\textbf{\textit{r}})$
\begin{equation}
E_{\mathrm{c}}^{\mathrm{nl}}= \frac{1}{2} \int \int
  d\textbf{\textit{r}} d\textbf{\textit{r'}}
  n(\textbf{\textit{r}}) \phi(\textbf{\textit{r}},\textbf{\textit{r'}})
  n(\textbf{\textit{r'}}).
\label{eq:NonLocal}
\end{equation}
where the kernel function $\phi(\textbf{\textit{r}},\textbf{\textit{r'}})$ is
discussed in detail in Ref.~\onlinecite{PRL92_246401}. Note that in our study
we employed both PBE and revPBE flavours of the GGA exchange-correlation
energy functional.

From Table~\ref{tab:ads_ene} it becomes apparent that in the case of the
perpendicular adsorption geometries, the vdW-DF values calculated with PBE
are in excellent agreement with those obtained by using the semi-empirical
approach DFT-D. This observation leads us to the preliminary conclusion that,
in the case of the covalently bonded systems, the DFT-D and vdW-DF will provide
similar results based on the PBE functional. However, when using the revPBE,
the adsorption energies calculated with the vdW-DF approach are higher
by $\approx$0.3 eV that those obtained with DFT-D method based on PBE. This
behaviour is similar to that observed for chemisorption energy of atoms and
molecules on transition-metal surfaces as pointed out by Hammer
\textit{et al.}\cite{PRB59_7413}

\subsubsection{Parallel adsorption geometry}

\paragraph*{Role of the vdW interactions} --
In our study we also focused on the thymine-Cu(110) surface interaction when
the molecule lies flat on the substrate. More specifically, we considered two
different adsorption configurations such that the molecule is oriented with
the line given by O--N--O along $[001]$ direction ($[001]$ geometry) or along
$[1\bar{1}0]$ one ($[1\bar{1}0]$ geometry). Intuitively, in these adsorption
configurations the molecule-surface bonding mechanism implies an interaction
between a $\pi$ molecular orbital and surface states which eventually leads to
a weakly bonded system. As was already discussed in literature for benzene and
naphthalene molecules on graphite(0001),\cite{PRL96_146107} phenol on
graphite(0001) and $\alpha$-Al$_2$O$_3$(0001)\cite{PRB74_155402} or for
thiophene on Cu(110) surfaces,\cite{PRL99_176401} the van der Waals
correlations play a crucial role to correctly describe the energetics of such
systems.

Indeed, the PBE calculations performed for both flat geometries predict an
average molecule-substrate distance of $\approx$3.4\,{\AA} that is a clear
fingerprint of a physisorption process. However, by including the dispersion
effects and relaxing the adsorbate-substrate interface, in both configurations
the molecule-surface distance drops to $\approx$2.7\,{\AA}. Besides this, the
thymine becomes tilted with respect to the surface by an angle of
$\approx$111$^\circ$ and $\approx$104$^\circ$ for $[001]$ and $[1\bar{1}0]$
geometries, respectively. This behavior is similar to that observed for
pyridine on Cu(110) and Ag(110) surfaces\cite{PRB78_045411} and clearly
emphasizes the importance of taking into account the structural relaxations
for such weakly bonded molecule-surface systems. From bonding point of view,
the main difference with respect to the perpendicular adsorption configurations
is that a flat thymine molecule is anchored to Cu(110) via a O--Cu bond with
a bondlength of $\approx$2.1\,{\AA}.

The necessity to include the dispersion corrections when investigating van der
Waals systems is illustrated by the analysis of the calculated adsorption
energies. As shown in Table~\ref{tab:ads_ene}, the PBE calculations predict
actually a \textit{non-bonding} thymine-surface configuration since the
adsorption energies are positive. However, when considering the effect of the
van der Waals interactions on the semi-empirical level, these energies are
lowered on average by $\approx$0.5 eV and become negative, denoting that a
flat adsorption geometry of thymine on Cu(110) substrate is possible. In this
case the $[001]$ adsorption geometry is more stable than the $[1\bar{1}0]$ one.

\begin{figure}
\begin{center}
  {\includegraphics[width=1.0\linewidth]{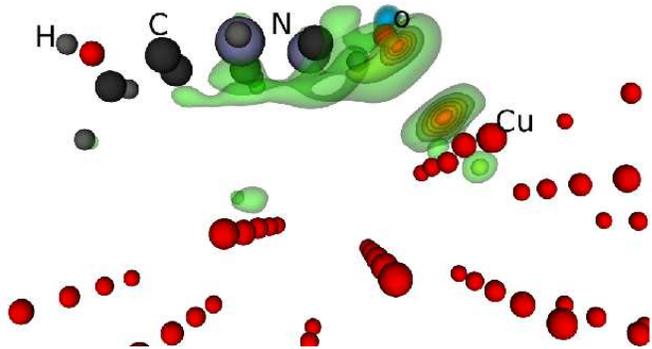}} \\
\end{center}
\caption{(Color online) Perspective 3D-view of the the 
nonlocal correlation energy difference
$\Delta E_{c}^{nl}$ of the thymine-Cu(110) surface system with respect to its
isolated surface and molecule components. The red color shows regions with a
large contribution to the binding energy while the green color represents
regions with a much smaller contribution.
}
\label{fig:EneDiff}
\end{figure}

The use of the vdW-DF method does not qualitatively change the physical picture 
highlighted above. From quantitative point of view, the vdW-DF adsorption
energies obtained with the PBE functional are lower by a factor of $\approx$2
with respect to those evaluated with the DFT-D approach. Similarly to the
case of the perpendicular thymine-surface geometries, the vdW-DF adsorption
energies calculated with the revPBE are about $\approx$0.5 eV higher then their
PBE counterparts. Although the relative stability of the flat adsorption
geometries considered in our study is not changed with respect to the PBE or
the semi-empirical method, the use of revPBE leads to a very small value of the
adsorption energy in the case of the $[1\bar{1}0]$ configuration. This is
consistent with the general tendency of the revPBE functional to provide 
theoretical chemisorption energies higher than the PBE one and thus presumably
closer to the corresponding experimental values.\cite{PRB59_7413} However,
note that the revPBE does not generally lead to an overall better physical
description than the PBE functional. For instance, in the case of a typical
van der Waals system such as Kr dimer\cite{JPCM19_305004} while the depth of
the Kr-Kr interaction potential is better described by revPBE, the equilibrium
distance of this dimer is better reproduced by the PBE functional.

Since we applied the vdW-DF method to an initially weakly bonded system that
in the final relaxed configuration exhibits a chemical bond, it is interesting
to determine the role of the nonlocal correlation effects in this case.
Therefore in Fig.~\ref{fig:EneDiff} we present a plot of the difference in the
nonlocal correlation energy due to the adsorption process, i.e., we map
in real space the quantity
$\Delta E_{c}^{nl} = E_{c}^{nl,sys}-E_{c}^{nl,Cu(110)}-E_{c}^{nl,thymine}$.
The peculiar feature displayed by this image is a strong contribution of the
nonlocal effects to the bonding energy arising from the charge density
$n(\textbf{\textit{r}})$ close to the O and Cu atoms that form a chemical bond.
Qualitatively this behavior can be understood by noticing that to form a
chemical bond between the O and Cu atoms, the charge density will shift from
these atoms to the space between them. Therefore, they become polarized which
favors a significant contribution of the nonlocal correlation energy given
by Eq.~\ref{eq:NonLocal} to the binding energy of the thymine-Cu(110) system.

\begin{figure}[tb]
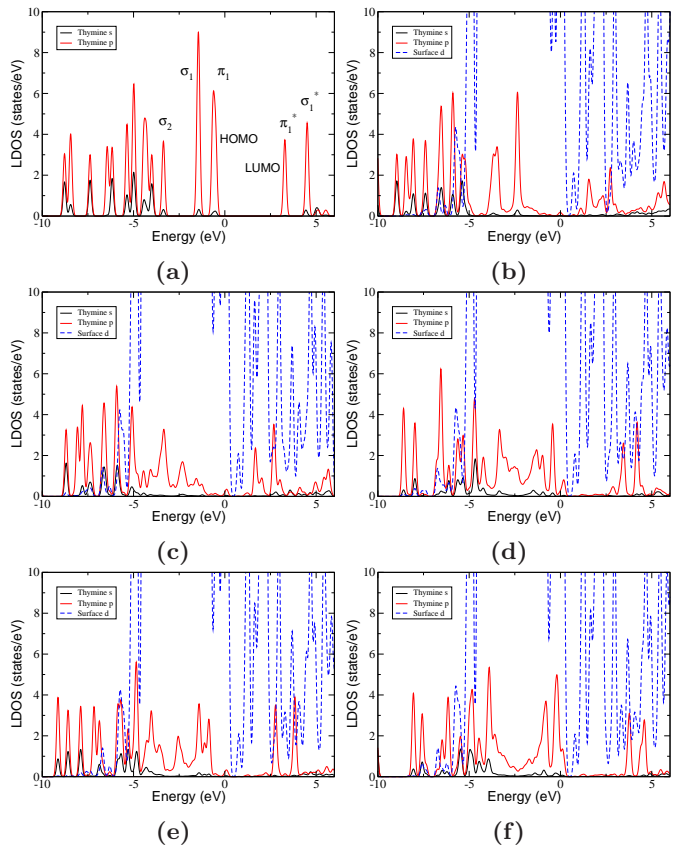

\begin{center}
  \begin{tabular}{cc}
  {\includegraphics[width=0.5\linewidth]{Thymine}}    &
  {\includegraphics[width=0.5\linewidth]{flat_001}} \\
  {\textbf{(a)}} & {\textbf{(b)}} \\
  {\includegraphics[width=0.5\linewidth]{ONO_H_geom}} &
  {\includegraphics[width=0.5\linewidth]{ONO_H2_geom}} \\
  {\textbf{(c)}} & {\textbf{(d)}} \\
  {\includegraphics[width=0.5\linewidth]{NO_H_geom}}  &
  {\includegraphics[width=0.5\linewidth]{NO_H2_geom}}  \\
  {\textbf{(e)}} & {\textbf{(f)}} \\
  \end{tabular}
\end{center}
\caption{
Local density of states (LDOS) calculated for \textbf{(a)} the thymine in the
gas phase, for \textbf{(b)} a flat thymine-surface adsorption geometry [see 
Figs.~\ref{fig:Thym}(d) and (e)]and for the \textbf{(c)} ONO-H, \textbf{(d)}
ONO-H$_2$, \textbf{(e)} NO-H and \textbf{(f)} NO-H$_2$ perpendicular adsorption
geometries of the C$_5$H$_6$N$_2$O$_2$ molecule on Cu(110) surface.
}
\label{fig:LDOS}
\end{figure}
\begin{figure}[tb]
\begin{center}
  \begin{tabular}{cc}
  {\textbf{+1.9 eV}} & {\textbf{-1.9 eV}} \\
  {\includegraphics[width=0.3\linewidth]{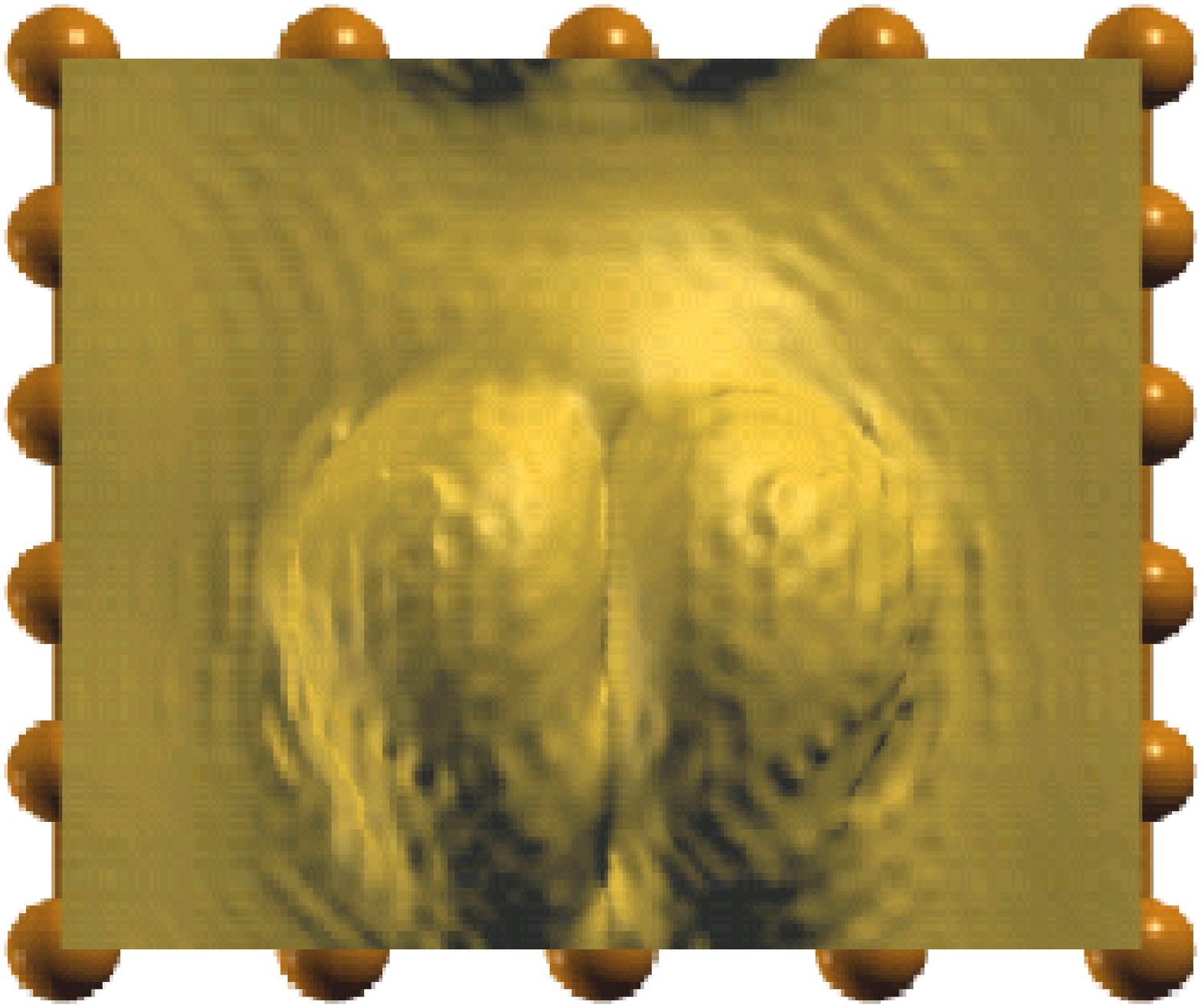}} &
  {\includegraphics[width=0.3\linewidth]{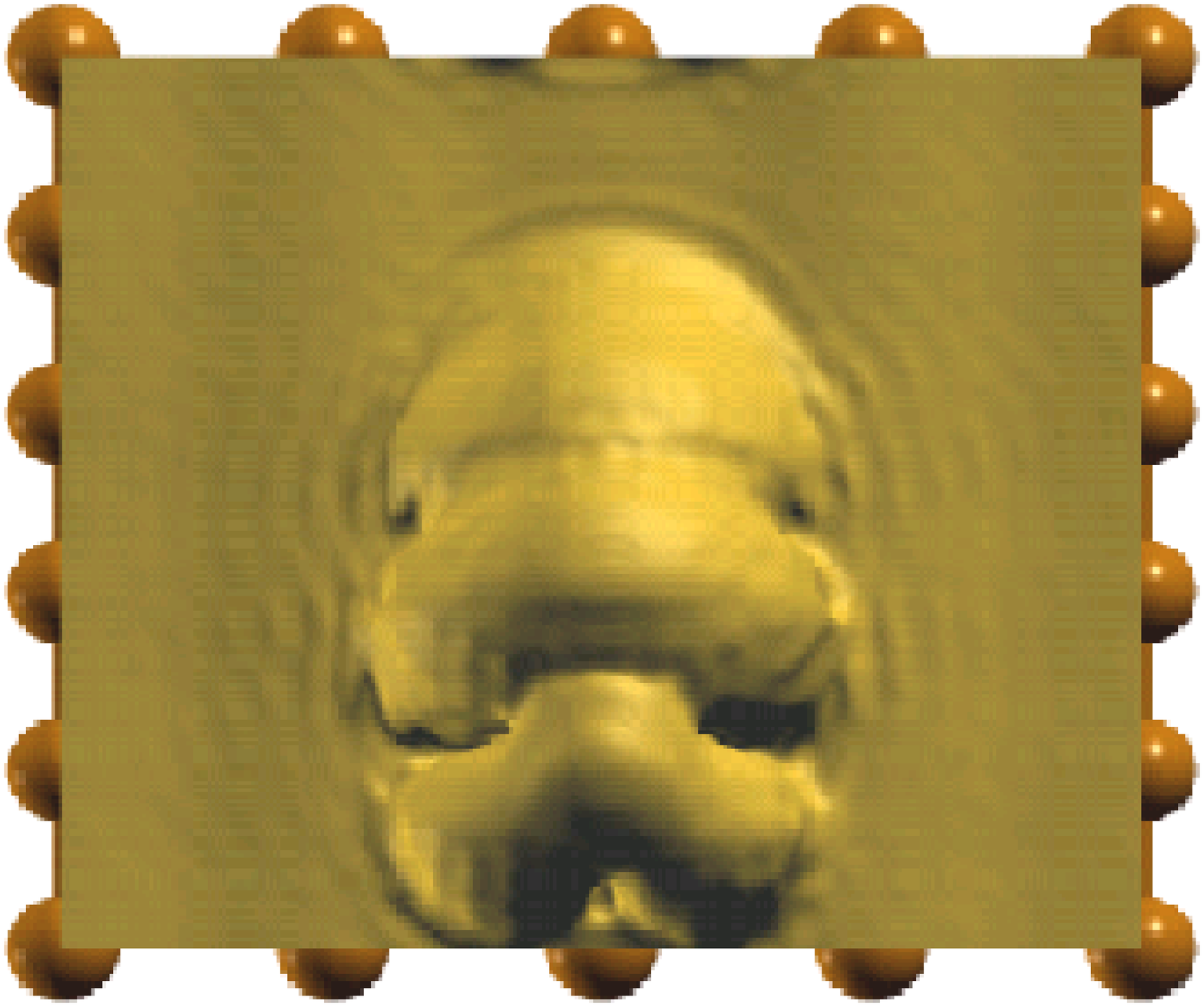}} \\
  {\textbf{(a)}} & {\textbf{(b)}} \\
  {\includegraphics[width=0.3\linewidth]{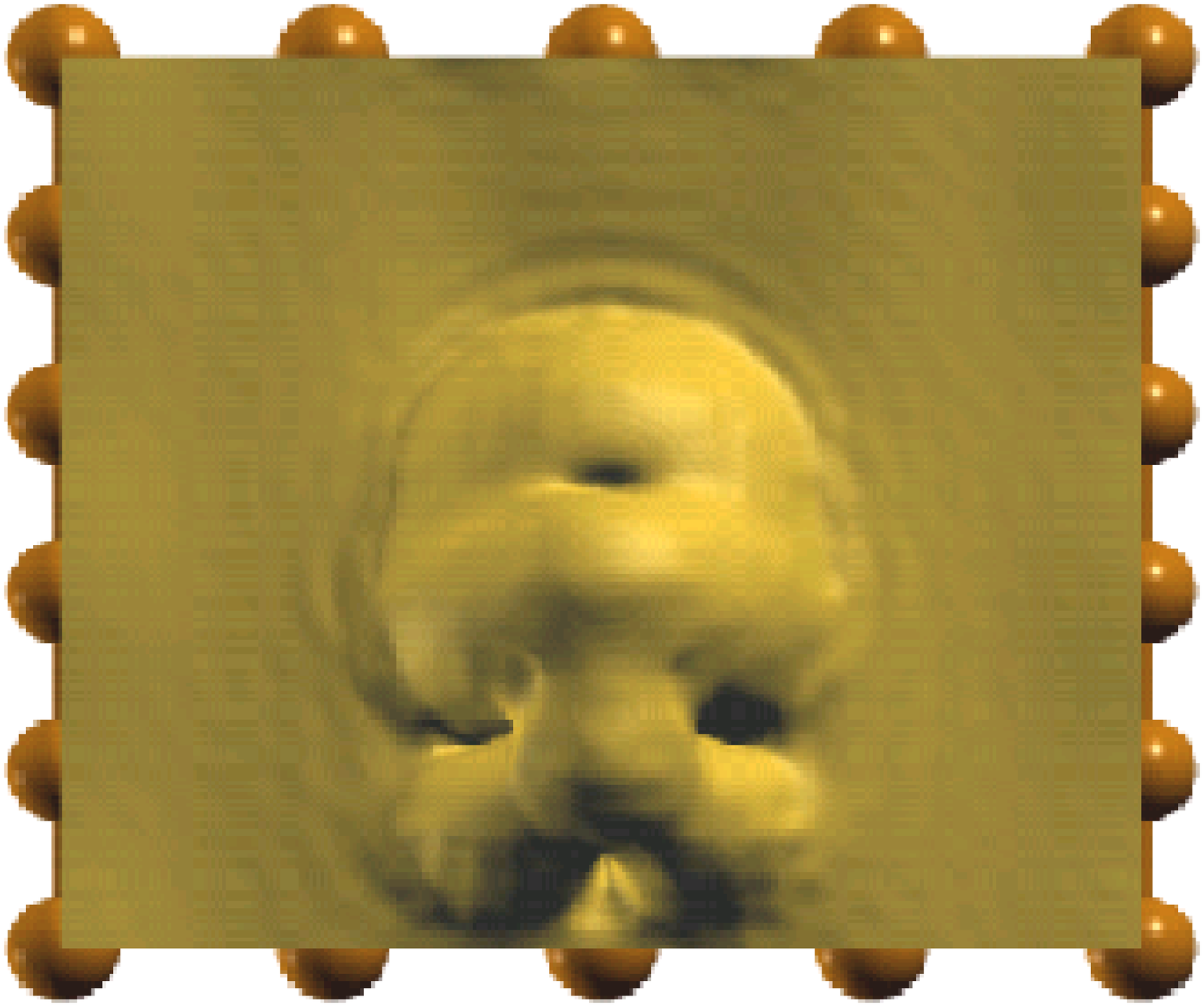}} &
  {\includegraphics[width=0.3\linewidth]{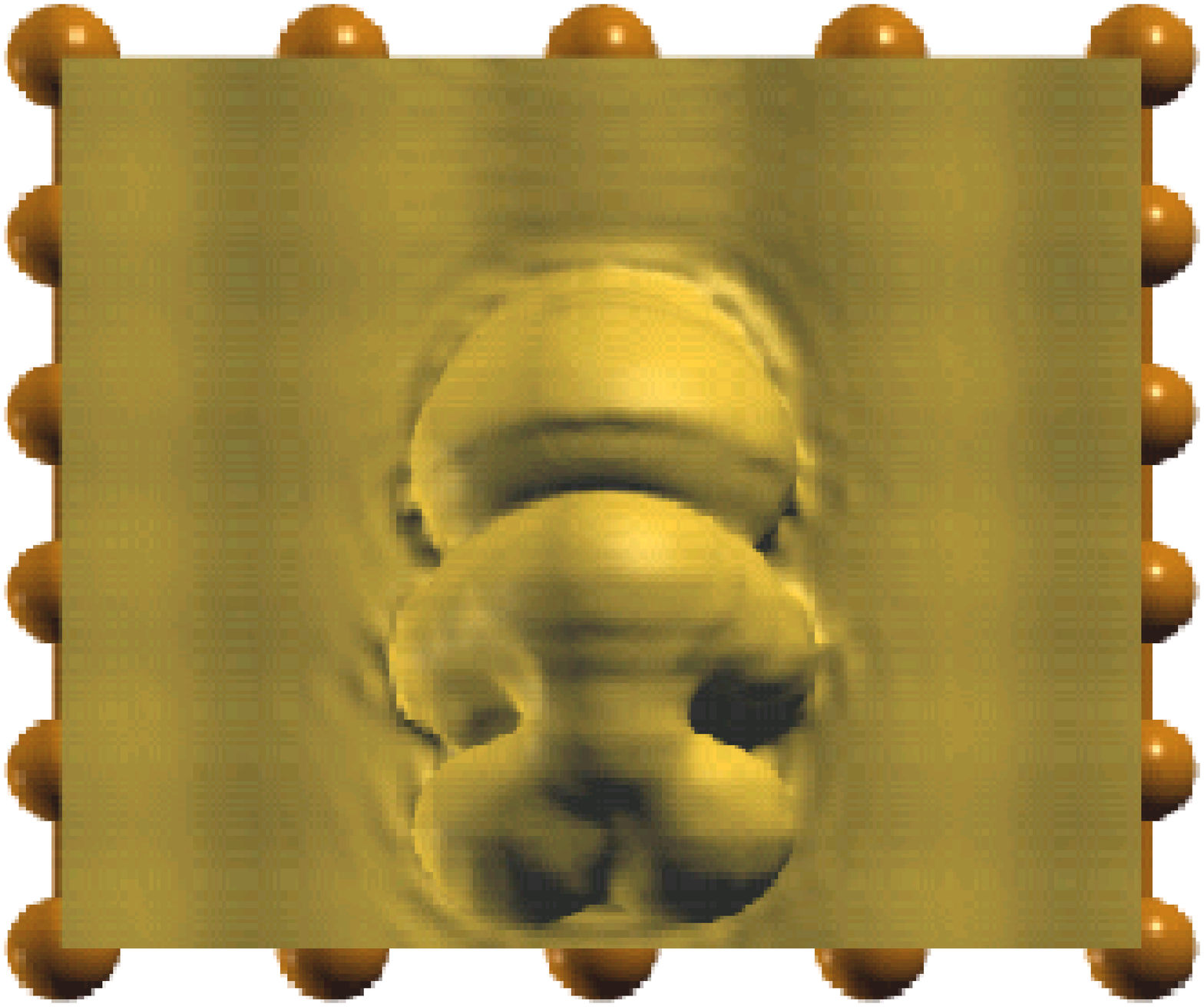}} \\
  {\textbf{(c)}} & {\textbf{(d)}} \\
  {\includegraphics[width=0.3\linewidth]{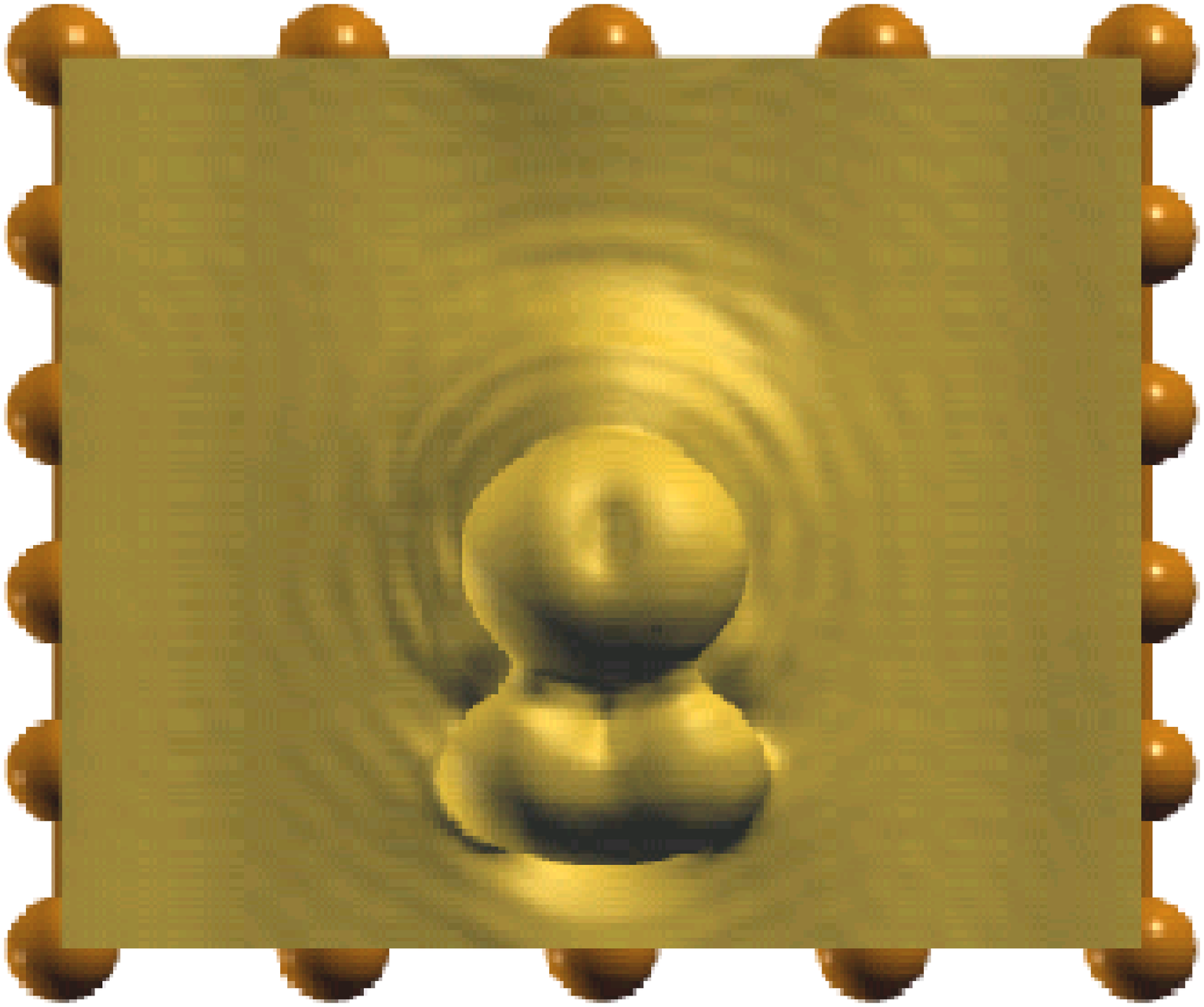}} &
  {\includegraphics[width=0.3\linewidth]{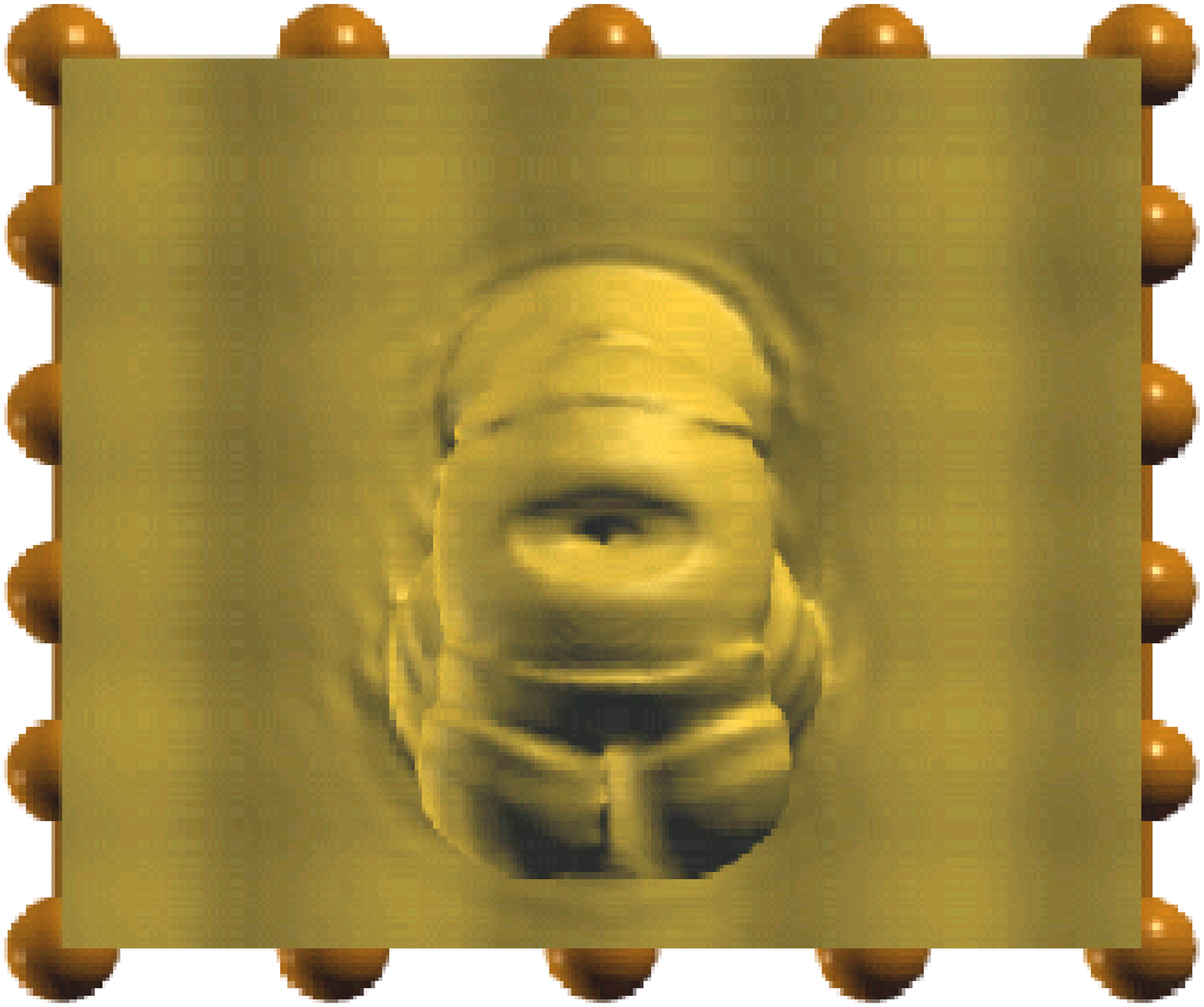}} \\
  {\textbf{(e)}} & {\textbf{(f)}} \\
  {\includegraphics[width=0.3\linewidth]{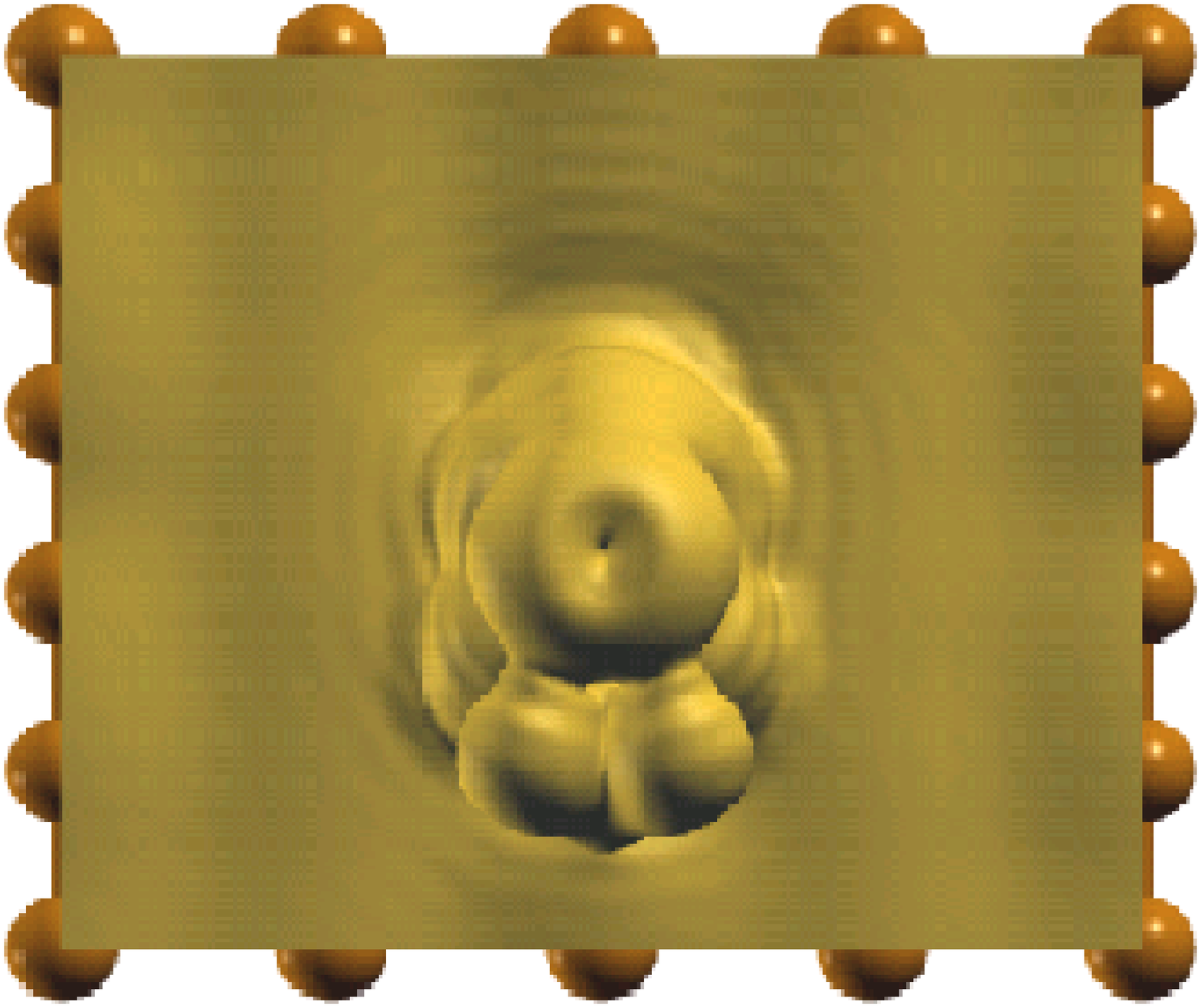}} &
  {\includegraphics[width=0.3\linewidth]{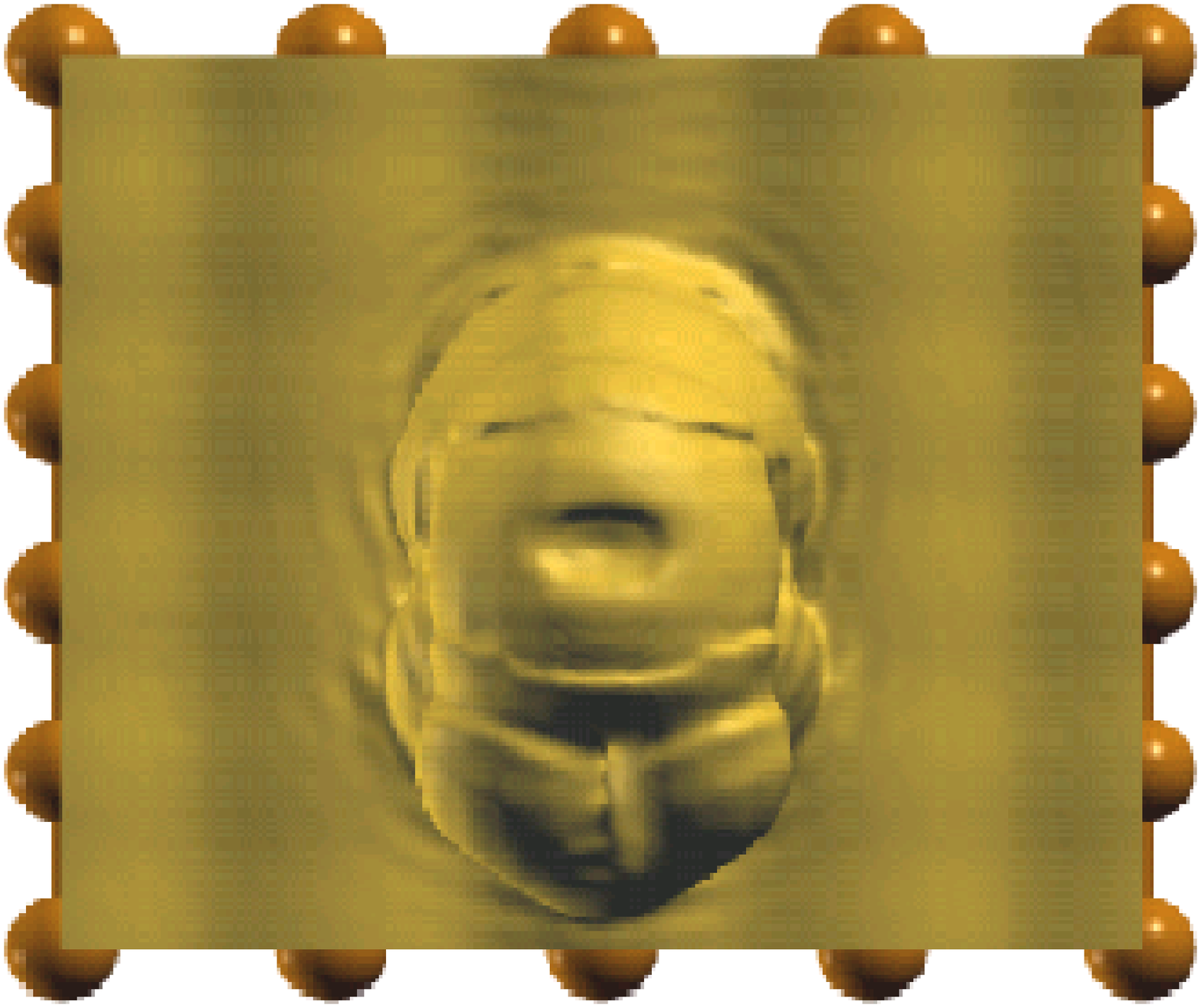}} \\
  {\textbf{(g)}} & {\textbf{(h)}} \\
  {\includegraphics[width=0.3\linewidth]{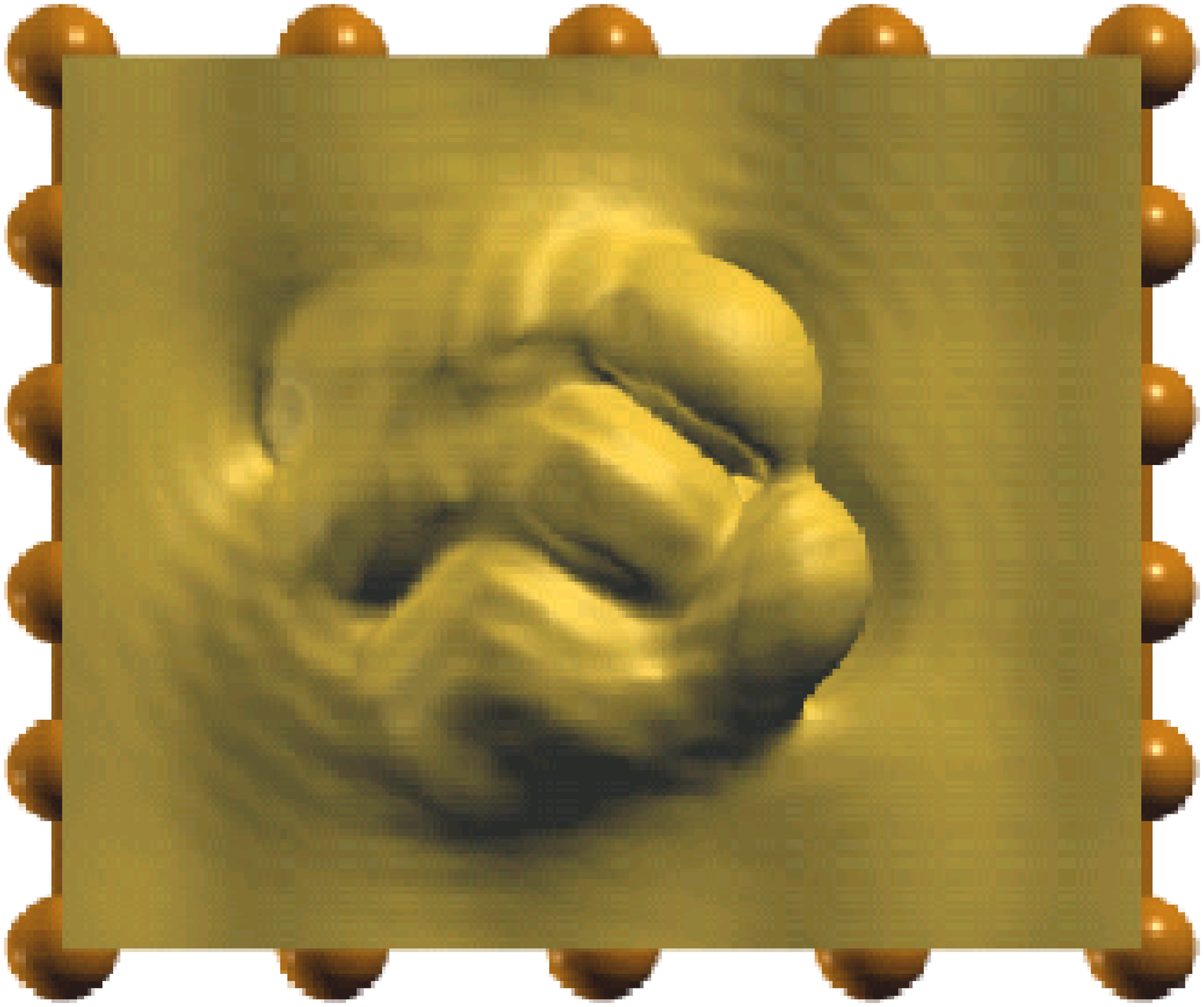}} &
  {\includegraphics[width=0.3\linewidth]{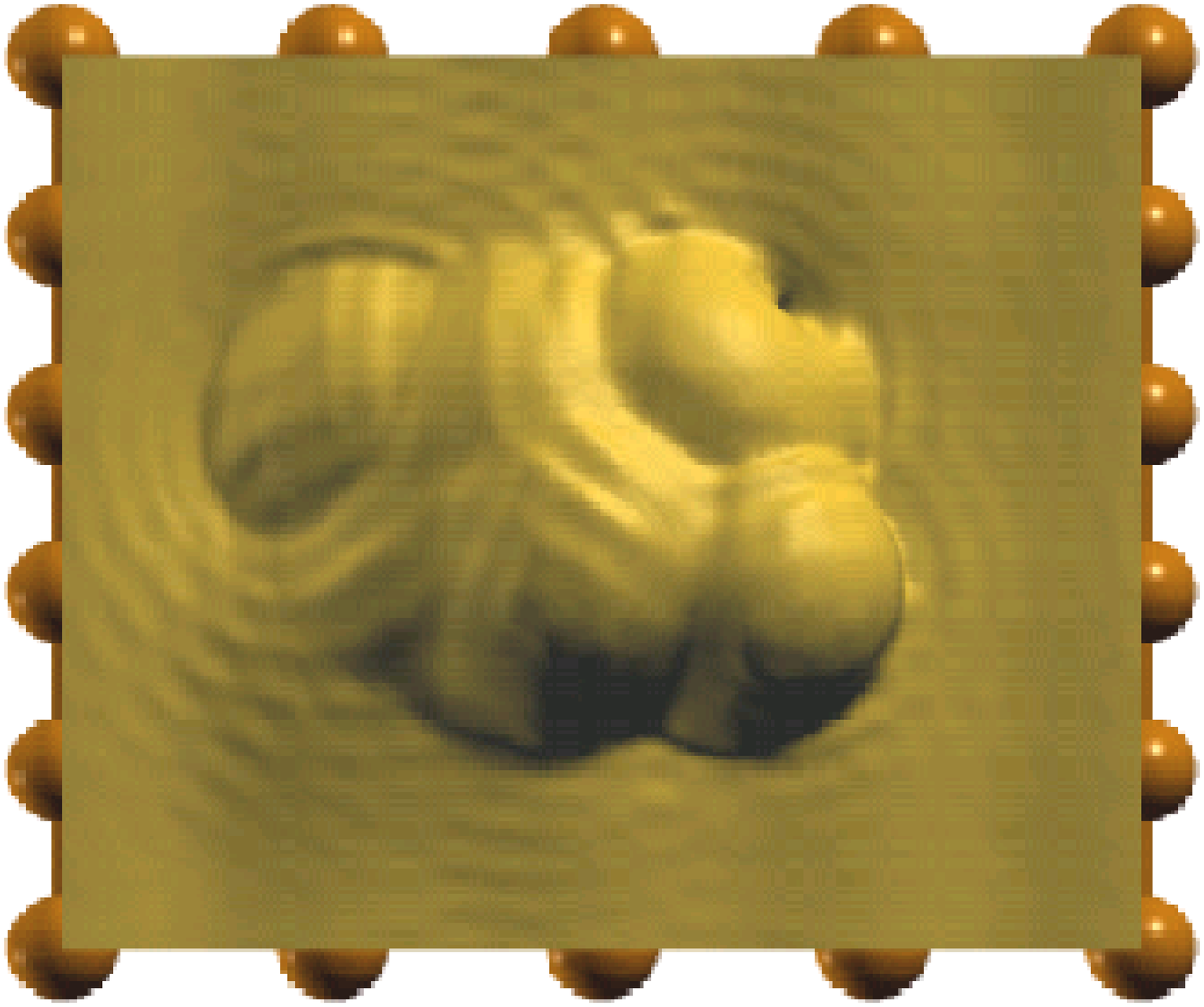}} \\
  {\textbf{(i)}} & {\textbf{(j)}} \\
  \end{tabular}
\end{center}
\caption{The simulated STM topographies for \textbf{(a,b)} ONO--H,
\textbf{(c,d)} ONO--$H_2$, \textbf{(e,f)} NO--H  and \textbf{(g,h)} NO--$H_2$
perpendicular adsorption configurations. The similar images for a flat
adsorption geometry [see Fig.~\ref{fig:Thym}(d) and (e)] are shown in
\textbf{(i)} and \textbf{(j)}. The applied bias voltages are +1.9 and -1.9 eV,
respectively. At a positive applied bias voltage the electrons will tunnel
from the STM tip to the unoccupied electronic states of the molecule-surface
interface.
}
\label{fig:STM}
\end{figure}
\subsection{Electronic structure}
\label{sec:elec}

The calculated atom-projected local density of states (LDOS) for all adsorption
configurations are presented in Fig.~\ref{fig:LDOS}. A general feature of the
electronic structure of all perpendicular adsorption geometries considered in
our study is a strong hybridization of the highest occupied molecular orbitals
(HOMOs) with the $d$-states of the Cu(110) surface in an energy range
of 5 eV below the Fermi level. This strong molecule-surface interaction leads
to broad bands with a mixed molecular-metal character. Besides this, in the
energy range from -5 to -10 eV, the interaction of the molecular states with
$s$- and $p$-states of the substrate become also important. As regarding the
most stable flat adsorption geometry depicted in Fig.~\ref{fig:Thym}(d), its
electronic structure also involves mainly a strong hybridization of HOMOs with
the $d$-states of the substrate.

However, in an energy range of
$\pm$2 eV around Fermi level, the LDOS has specific features for each
studied adsorption configuration. This observation is important since the
electronic structure of adsorbate-metal surface in this energy range can be
probed by using the scanning tunneling microscope (STM). Therefore, using the
Tersoff and Hamann theory,\cite{PRB31_805} we simulated maps of constant
current for the ground-state perpendicular and parallel adsorption geometries
considered in our study. The theoretical STM topographies are depicted in
Fig.~\ref{fig:STM} for two different applied bias voltages. In general, the
simulated STM images exhibit peculiar features for each adsorbate-surface
configuration at each value of the bias voltage except for the NO--H and
NO--H$_2$ geometries which exhibit similar features.

Among the simulated theoretical STM images, those obtained for the ONO--H
geometry at a bias voltage of $\pm$1.9 eV present the most significant
topographic difference. Also, except for this configuration, the simulated
images suggest the possibility to obtain a sub-molecular resolution for single
or double deprotonated thymine molecule adsorbed in a perpendicular
configuration on Cu(110) surface. For instance, the methyl group (CH$_3$)
appears as a distinct clear feature, especially in the case of the ONO--H$_2$
and NO--H geometries although its geometrical form is specific for each
configuration in question. A similar sub-molecular resolution of the STM
images is observed for a parallel adsorption geometry corresponding to the
configuration shown in Fig.~\ref{fig:Thym}(d).
\section{\label{sec:concl} Summary}

By using the \textit{ab initio} pseudopotential method, we analyzed the
adsorption process of a single thymine molecule on the Cu(110) surface. To
determine the ground-state adsorption geometry, we investigated several
adsorbate-substrate configurations such that the molecular plane is
perpendicular or parallel to the (110) surface. To explore the possibility to
chemically functionalize this molecule-surface system, for the perpendicular
adsorption geometry we considered the case of a single and double deprotonated
thymine molecule anchored to surface via different functional groups. Our
\textit{ab initio} simulations indicate that the strongest thymine-surface
interaction takes place when the heterocycle ring is perpendicular to substrate
and aligned along its $[1\bar{1}0]$ direction. For this ground-state adsorption
geometry, the bonding of thymine on Cu(110) surface is due to a strong
hybridization of the highest occupied molecular orbitals (HOMOs) with the
$d$-states of the substrate. In the case of a flat adsorption geometry, our
calculations demonstrate that the long-range van der Waals interactions play a
very important role from both geometrical and energetic point of views. The
effect of the dispersion corrections on the geometry of the relaxed
adsorbate-surface interface was determined by using a semi-empirical approach.
If the relaxed geometry obtained from PBE calculations remains almost parallel
to surface, the inclusion of the van der Waals interactions leads to tilted
adsorption geometry by an angle of 104$^\circ$. Moreover, the van der Waals
corrections have a more dramatic impact on the calculated adsorption energies.
The PBE adsorption energies obtained for the flat adsorption geometries
are positive and thus the PBE calculations suggest a \textit{non-bonding}
molecule-surface configuration. On the contrary, the adsorption energies
calculated with the semi-empirical and the vdW-DF methods are negative and
thus the parallel adsorption configuration becomes \textit{bonding}. In this
case the inclusion of the van der Waals interactions significantly modifies
the character of the adsorption process from physisorption to chemisorption.
As a final remark, we also simulated the scanning tunneling microscopy (STM)
topography of the thymine-Cu(110) surface that can serve to distinguish in
experiments between various adsorption configurations of the thymine-Cu(110)
interface.

\section*{Acknowledgments}

The computations were performed with the help of the JUMP and Blue/Gene
supercomputers at the Forschungszentrum J\"ulich, Germany. We acknowledge the
financial support from the DFG (Grants No. SPP1243 and No. HO 2237/3-1), 
Alexander von Humboldt foundation and Japan Society for the Promotion of
Science. V.C., N.A. and P.L. thank to H. H\"olscher, J.-H. Franke, H. Fuchs
(University of M\"unster), K. Schroeder (Research Center J\"ulich) and R. Brako
(Rudjer Bo\v skovi\'c Institute) for many fruitful discussions and continuing
support.

\section*{Appendix}
A detailed description of how to evaluate the surface free energies using the
total energies obtained from DFT calculations can be found in 
Refs.~\onlinecite{PRB44_1419} and~\onlinecite{PRB65_035406}. In the following
we will derive the equations used in the present work to calculate the
relative surface free energies $\Delta \gamma$ plotted in
Fig.~\ref{fig:SurfEne}.

For a deprotonation process characterized by the relative Gibbs free energy
$\Delta G^{ads}$, the relative surface free energy ($\Delta \gamma$) can be
written as 
\begin{align}
\Delta\gamma = (1/A) * (\Delta G^{ads} + factor * \mu_{H}).
\nonumber
\end{align}
where $A$ represents the surface area of the unit cell and $\mu_{H_{2}}$ is the 
chemical potential of the $H_2$ molecule. The $factor$ is equal 0.0 for thymine 
molecule, 1.0 for a single deprotonated or to 2.0 for a double deprotonated
thymine molecule. 

Assuming that the entropic term is small as compared to the values of the
adsorption entalpies at $T=0$~K,\cite{PRB65_035406} we approximate
$\Delta G^{ads}$ by $\Delta H_{T=0}^{ads}$. Therefore, the relative surface
free energies $\Delta \gamma$ of the ground-state flat, ONO--H and ONO--H$_2$
configurations are given by:
\begin{align}
\begin{split}
A* \Delta\gamma_{_{flat}}  &\approx \Delta H_{T=0}^{ads}   + 0.0*\mu_{H}
                           = -0.260 + 0.0 * \mu_{H}, \\
A* \Delta\gamma_{_{ONO-H}} &\approx \Delta H_{T=0}^{ads}   + 1.0*\mu_{H}
                           = -0.872 + 1.0 * \mu_{H}, \\
A* \Delta\gamma_{_{ONO-H_2}} &\approx \Delta H_{T=0}^{ads} + 2.0*\mu_{H}
                           = +0.544 + 2.0 * \mu_{H}.
\end{split}
\nonumber
\end{align}
where $\mu_{H}$ has been calculated according to Eq.~(22) from the
Ref.~\onlinecite{PRB65_035406} and the values of $\Delta H_{T=0}^{ads}$
correspond to those obtained with the DFT-D method.
\bibliography{Thymine_literature}
\end{document}